\documentclass[letterpaper,10pt,journal,twoside]{IEEEtran}
\usepackage{listings}
\usepackage[auth-lg]{authblk}
\usepackage{algpseudocode}                                 % new algorithm package
\usepackage{algorithm}
\usepackage{graphicx}                                      % include pdf figures
\usepackage{amsmath}
\usepackage{amssymb}
\usepackage{amsfonts}
\usepackage{booktabs}
\usepackage{multirow}
\usepackage[subrefformat=parens,farskip=0pt,justification=centering]{subfig}
\usepackage{color}
\usepackage{cite}                                          % more citations in one bracket
\usepackage{comment}                                       % use comment
\usepackage{soul}                                          % use highlight command \hl{}
\soulregister\cite7
\soulregister\ref7
\soulregister\pageref7
\usepackage{amsthm}
\usepackage{etoolbox}                                      % commands \newtoggle, \toggletrue, \iftoggle
\usepackage{url}
\usepackage{hyperref}
\usepackage[hyphenbreaks]{breakurl}
\usepackage{nth}                                           % nth command
\usepackage{bm}                                            % bm command
\usepackage{balance}
\usepackage{threeparttable}
\usepackage{booktabs}
\usepackage{colortbl}
\usepackage{filecontents}                                  % support to pgfplots
\usepackage{pgfplots}
\usepackage{pgfplotstable}
\pgfplotsset{compat=newest}
\usepackage{caption}
\newcommand\VRule[1][\arrayrulewidth]{\vrule width #1}
\usepackage[normalem]{ulem}                                            % for underline

\algrenewcommand\textproc{\texttt}

\makeatletter
\let\OldStatex\Statex
\renewcommand{\Statex}[1][3]{%
  \setlength\@tempdima{\algorithmicindent}%
  \OldStatex\hskip\dimexpr#1\@tempdima\relax
}
\makeatother

\RequirePackage[normalem]{ulem} %DIF PREAMBLE
\RequirePackage{color}\definecolor{RED}{rgb}{1,0,0}\definecolor{BLUE}{rgb}{0,0,1} %DIF PREAMBLE
\lstdefinestyle{base}
{
    language   = C,
    emptylines = 1,
    breaklines = true,
    basicstyle = \ttfamily\scriptsize,
    moredelim  = **[is][\color{red}]{@}{@},
}

\begin{document}

\title{SPINBIS: \uline{Spin}tronics based \uline{B}ayesian \uline{I}nference \uline{S}ystem with Stochastic Computing}

\author{Xiaotao~Jia,~\IEEEmembership{Member,~IEEE,}
        Jianlei~Yang,~\IEEEmembership{Member,~IEEE,}
        Pengcheng~Dai,
        Runze~Liu,
        Yiran~Chen,~\IEEEmembership{Fellow,~IEEE}
        and~Weisheng~Zhao,~\IEEEmembership{Fellow,~IEEE}
\thanks{Manuscript received on May 2018, and revised on September 2018, October 2018 and December 2018, accepted on January 2019. This work was supported in part by the National Natural Science Foundation of China (61602022, 61501013, 61571023, 61521091 and 1157040329), State Key Laboratory of Software Development Environment (SKLSDE-2018ZX-07), National Key Technology Program of China (2017ZX01032101), CCF-Tencent IAGR20180101 and the 111 Talent Program B16001. \textit{Xiaotao Jia and Jialnei Yang contributed equally to this work. Corresponding authors are Jianlei Yang and Weisheng Zhao.}}
\thanks{X. Jia, P. Dai and W. Zhao are with Beijing Advanced Innovation Center for Big Data and Brain Computing, Fert Beijing Research Institute, School of microelectronics, Beihang University, Beijing, 100191, China. E-mail: weisheng.zhao@buaa.edu.cn}
\thanks{J. Yang and R. Liu are with Beijing Advanced Innovation Center for Big Data and Brain Computing, Fert Beijing Research Institute, School of Computer Science and Engineering, Beihang University, Beijing, 100191, China. E-mail: jianlei@buaa.edu.cn}
\thanks{Y. Chen is with Department of Electrical and Computer Engineering, Duke University, Durham, NC 27708, USA.}
}

\maketitle
\thispagestyle{empty}

\begin{abstract}

Bayesian inference is an effective approach for solving statistical learning problems, especially with uncertainty and incompleteness. However, Bayesian inference is a computing-intensive task whose efficiency is physically limited by the bottlenecks of conventional computing platforms. In this work, a spintronics based stochastic computing approach is proposed for efficient Bayesian inference. The inherent stochastic switching behaviors of spintronic devices are exploited to build stochastic bitstream generator (SBG) for stochastic computing with hybrid CMOS/MTJ circuits design. Aiming to improve the inference efficiency, an SBG sharing strategy is leveraged to reduce the required SBG array scale by integrating a switch network between SBG array and stochastic computing logic. A device-to-architecture level framework is proposed to evaluate the performance of spintronics based Bayesian inference system (SPINBIS). Experimental results on data fusion applications have shown that SPINBIS could improve the energy efficiency about $12\times$ than MTJ-based approach with $45\%$ design area overhead and about $26\times$ than FPGA-based approach.

\end{abstract}


\IEEEpeerreviewmaketitle

 \iffalse

 \bibliography{../ref/Bayesian}

 \fi

\section{Introduction} \label{Section:Intro}

\IEEEPARstart{T}{he} rise of deep learning has greatly promoted the development of artificial intelligence. However, most deep learning approaches usually require large scale training data and also bring some overfitting problems. Meanwhile, they can neither represent the uncertainty and incompleteness of the world nor take the advantages of well-studied human experience. In order to overcome these limitations, some researches trend to utilize Bayesian inference or combine Bayesian approaches with deep learning. Bayesian inference provides a powerful approach for information fusion, reasoning and decision making that has established it as the key tool for data-efficient learning, uncertainty quantification and robust model composition. It is widely used in applications of artificial intelligence and expert systems, such as multi-sensor fusion~\cite{pinheiro2004Bayesian} and Bayesian belief network~\cite{cruz2007diagnosis}. Recently Bayesian learning has drawn great attentions on deep learning community and is well combined with many deep neural networks~\cite{gal2017deep}.

\begin{figure}[tb!]
    \centering
    \includegraphics[scale=0.8]{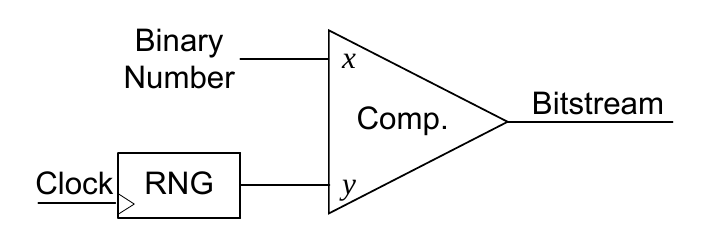}
    \caption{Traditional  stochastic bitstream generator (SBG) circuits. If $x>y$, then outputs `1'; otherwise outputs `0'.}
    \label{fig:sbg}
\end{figure}

The fundamental of Bayesian inference is Bayes' rule which could be implemented by probabilistic computing. Probability computing is a kind of computation-intensive task which is inefficient with deterministic computation mode~\cite{thakur2016Bayesian}. Stochastic computing (SC) is an unconventional computing mode which has observed to be suitable for efficient probability computing~\cite{alaghi2013survey} with high error tolerance abilities and low-cost implementations of arithmetic operations. However, it is difficult to leverage the parallelism of stochastic computing algorithms on traditional von-Neumann architectures~\cite{grollier2016spintronic}. Hence, reconfigurable approach~\cite{lin2010high} and analog computing~\cite{mroszczyk2014accuracy}\cite{friedman2016bayesian} is utilized to realize stochastic computing in order to improve the Bayesian inference efficiency. The stochastic computing are usually realized by bit-wise operations on stochastic bitstreams which is created by pseudo-random number generators (RNG) and comparators as shown in Fig.~\ref{fig:sbg}. It is still expensive to implement stochastic bitstream generator (SBG) on von-Neumann architectures with CMOS technologies which is critical for performing stochastic computing.

\begin{figure}[tb!]
    \centering
    \includegraphics{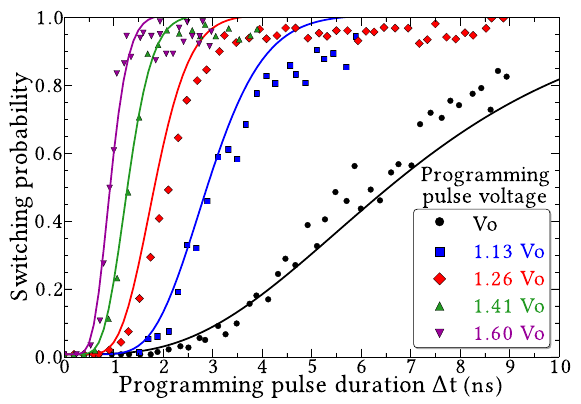}
    \caption{Experimental measurements of the switching probability with respect to the duration of the applied programming pulse, for different programming voltages \cite{vincent2015analytical}.}
    \label{fig:stochswitch}
\end{figure}

Recently, spintronic devices (such as magnetic tunnel junction, MTJ) pose some promising advantages on generating random numbers because of the stochastic switching behaviors~\cite{devolder2008single}. As shown in Fig.~\ref{fig:stochswitch}, an MTJ device usually switches with a nondeterministic manner according to the applied bias voltage and duration time due to the inherent thermal fluctuation of magnetization. Such a stochastic switching behavior has been exploited for generating random numbers~\cite{de2015stochastic,wang2016novel,wang2017hybrid} efficiently. And consequently, the inherent randomness of spintronic devices could be well revealed as the stochastic resources to perform stochastic computing.

In this paper, spintronic device based stochastic computing is proposed for efficient Bayesian inference system (SPINBIS). The main contributions of this work are listed as follows:
\begin{itemize}
    \item An efficient stochastic bitstream generator is proposed by leveraging the stochastic switching behaviors of MTJ device. The generated bitstreams have a very low correlation which is critical for stochastic computing accuracy. And a state-aware self-control strategy is adopted to improve the SBG efficiency.
    \item An SBG sharing strategy is leveraged to reduce the required SBG array scale by integrating a switch network between SBG array and stochastic computing logic. The power consumption of SPINBIS is greatly reduced benefiting from this strategy.
    \item A device-to-architecture level framework is built to evaluate the performance of SPINBIS with the data fusion applications. Experimental results indicate that it could achieve significant improvement on inference efficiency in terms of power, area and speed.
\end{itemize}

The remainder of this paper is organized as follows. Section~\ref{Section:Background} states some preliminaries and related works. The architecture of SPINBIS is presented in Section~\ref{Section:spinbis} as well as the SBG sharing techniques. Section~\ref{Section:sbg} describes the SBG circuit design and state-aware self-control strategy. A device-to-architecture evaluation framework and experimental results on typical applications are illustrated in Section \ref{Section:experiments}. Concluding remarks are given in Section~\ref{Section:Conclusion}.

 \iffalse

 \bibliography{../ref/Bayesian}

 \fi

\section {Background} \label{Section:Background}

\subsection{Stochastic Computing}

\begin{figure}[tb!]
    \centering
    \includegraphics[width=0.48\textwidth]{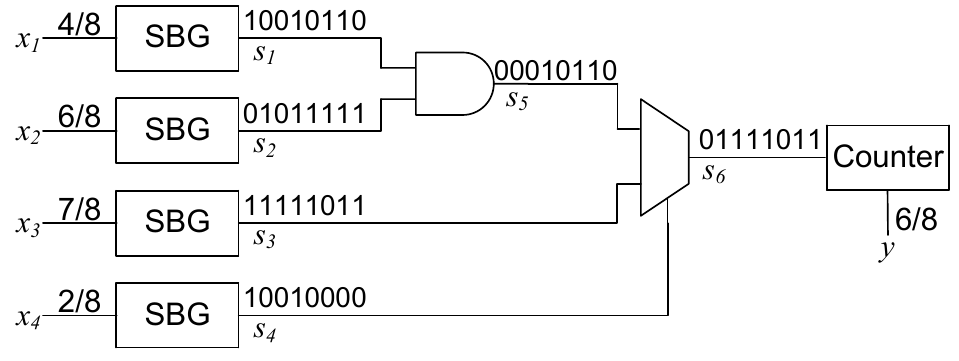}
    \caption{Stochastic circuit that realizes the arithmetic function $y=x_1x_2x_4+x_3(1-x_4)$.}
    \label{fig:sc}
\end{figure}

Stochastic computing was first introduced in the 1960's by von Neumann~\cite{von1956probabilistic}. The basic idea of stochastic computing is to represent probability value $p$ by the ratio of `1' in the binary bitstreams. It is obvious that the representation of $p$ by stochastic bitstream is not unique. The value of the bitstream is only related to its length and the count of `1', but has nothing to do with the position of `1'. There are two encoding formats for stochastic bitstream: unipolar and bipolar format. The value range of unipolar is $[0, 1]$ while bipolar is $[-1, 1]$. If the bitstream length is $n$, out of which $k$ bits are `1's, then the probability value $p$ is represented by $p = k / n$ if using unipolar encoding format or $p = (2k-n)/n$ if using bipolar encoding format. In this work, unipolar encoding format is adopted because $p \in [0,1]$ in Bayesian inference problem. Arithmetic operations in stochastic computing are realized by using simple logic gates. For example, the multiplication operation is achieved by an \texttt{AND} gate and scaled addition is achieved by a \texttt{MUX} as shown in Fig.~\ref{fig:sc}. Even though there exists a slight loss in computation accuracy, the advantage of stochastic computing is that it could significantly improve the computation energy efficiency when compared with conventional methods~\cite{venkatesan2015spintastic,mondal2017power,friedman2017approximation,calvet2016sleep,liu2018energy,liu2018stochastic}. It is very suitable for inherent error-resilience applications using stochastic computing to make a trade-off between the accuracy constraints and the energy efficiency requirements~\cite{ren2017sc,kim2016dynamic,li2017towards}.

Stochastic computing is not an exact computing method while the accuracy problem is arisen from several reasons. The first reason is that the probability values $p$ are usually converted to stochastic bitstream with a lower quantization accuracy compared with fixed or floating point methods. The second reason is that the correlations between different bitstreams usually degrade the computation accuracy since these bitstreams are usually obtained by pseudo random number generators. Aiming to improve the quality of SBGs, many pioneer researchers have proposed several SBG models such as linear feedback shift registers (LFSRs)~\cite{jeavons1994generating,cai2018vibnn,kim2016energy}, weighted binary SNG~\cite{gupta1988binary}. However, such CMOS based approaches usually pose some bottlenecks on power consumption and chip area efficiency. And consequently some emerging devices based approaches are investigated in this work.

\vspace{-2mm}
\subsection {Magnetic Tunnel Junction (MTJ) Device}

\begin{figure}[tb!]
    \centering
    \includegraphics[scale=0.8]{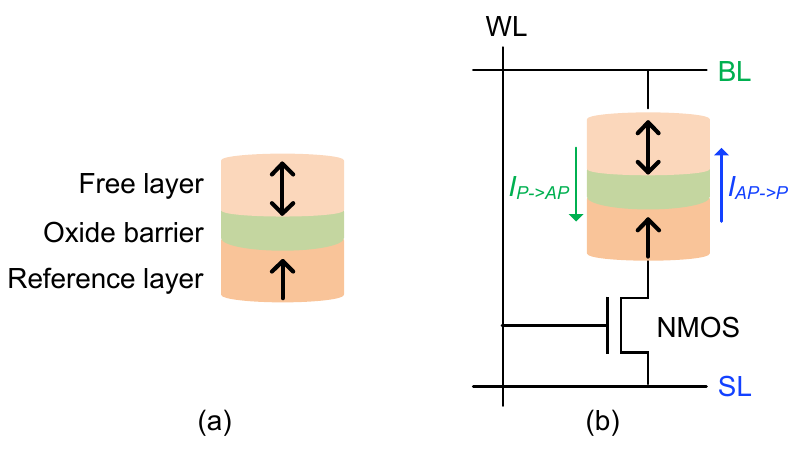}
    \caption{(a) The typical structure of PMA-MTJ. (b) The circuit schematic view of 1T1MTJ structure.}
    \label{fig:mtj}
\end{figure}

Fig.~\ref{fig:mtj}(a) shows a typical structure of the MTJ device with perpendicular magnetic anisotropy (PMA)~\cite{wang2018current}. MTJ is a sandwich structure consisting of two ferromagnetic (FM) layers and a tunneling barrier layer. One FM layer is defined as reference layer (PL) with fixed magnetization direction. Another FM layer is a kind of free layer (FL) whose magnetization direction could be parallel or anti-parallel with that of PL. The MTJ resistance is determined by the relative magnetization directions of PL and FL while parallel (\texttt{P}) magnetization behaviors as low-resistance ($R_P$) state (logic `0') and anti-parallel (\texttt{AP}) magnetization behaviors as high-resistance ($R_{AP}$) state (logic `1'). Tunnel magneto resistance ratio $TMR = (R_{AP}-R_{P})/R_{P}$ is defined to characterize the relationship of $R_P$ and $R_{AP}$. Fig.~\ref{fig:mtj}(a) shows the circuit schematic view of a popular \texttt{1T1MTJ} memory cell. MTJ state can be flipped by applying a polarized current injection with spin transfer torque (STT) mechanism. The switching current is controlled by the voltage between bit-line (BL) and the source-line (SL). The nMOS transistor serves as a switch and controlled by word-line (WL). As shown in Fig.~\ref{fig:mtj}(b), the MTJ state is switched from \texttt{P} state to \texttt{AP} state if the injected current ($I_{P \to AP}$) flows through the MTJ from FL to PL. On the contrary, the MTJ state is switched from \texttt{AP} state to \texttt{P} state if $I_{AP \to P}$ is injected. The MTJ state could be flipped only if the applied bias voltage is larger than a critical current $I_{c0}$ with an enough duration time as shown in Fig.~\ref{fig:stochswitch}.

The stochastic behavior of MTJ switching has been revealed by~\cite{devolder2008single} and is resulted from the unavoidable thermal fluctuations of magnetization~\cite{marins2012precessional}. The MTJ device usually switches with a stochastic manner according to the applied voltage magnitude and duration time as shown in Fig.~\ref{fig:stochswitch}, which could be represented as a random event with the probability $p$. Such stochastic behaviors have been evaluated in MRAM designs~\cite{zhang2013multi}, neural network~\cite{srinivasan2017magnetic}\cite{angizi2018leveraging} and random number generator~\cite{fong2014generating}. Meanwhile, such an inherent probabilistic switching property is a very promising approach to generate stochastic bitstreams for stochastic computing.

Recently, a simple MTJ based SBG is proposed in~\cite{de2015stochastic} but it lacks of many circuit details. In~\cite{onizawa2016analog}, an MTJ based analog-to-stochastic converter is proposed for stochastic computation in vision chips. In~\cite{mondal2017power}, MTJ based stochastic computing is integrated into artificial neural network applications. However, the energy efficiency of their SBGs is relative low. Furthermore, they have not considered the correlation between different SBGs which will significantly degrade the computation accuracy. Voltage-controlled MTJs (VC-MTJs) are introduced for stochastic computing to reduce the power consumption in~\cite{wang2017hybrid} but each SBG involves too many MTJs. Bitstream correlation is discussed in~\cite{wang2017hybrid}, however, the proposed shuffle operation could not remove the relevance essentially and may still result in unexpected results. 
 \iffalse

 \bibliography{../ref/Bayesian}

 \fi

\section {The SPINBIS architecture} \label{Section:spinbis}

\subsection{Motivation} \label{Section:spinbis:motivation}
\begin{figure}[tb!]
    \centering
    \includegraphics[width=0.45\textwidth]{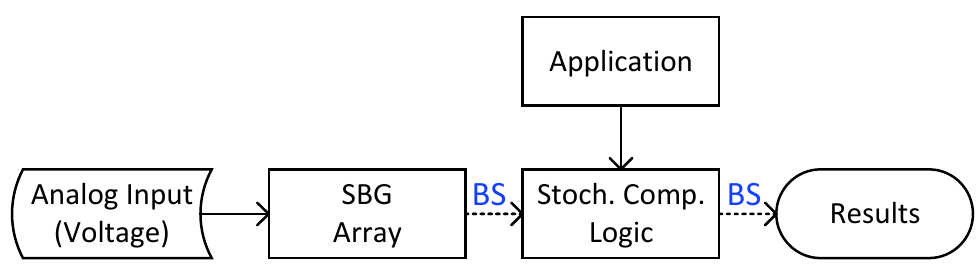}
    \caption{A typical Bayesian inference system \cite{jia2017spintronics}.}
    \label{fig:aspdacarchi}
\end{figure}

A typical Bayesian inference system (BIS)~\cite{jia2017spintronics} is shown in Fig.~\ref{fig:aspdacarchi}. The input of BIS is a set of bias voltages corresponding to evidence or likelihood. SBG array is utilized to generate stochastic bitstreams (SB) according to the input voltages. The bitstreams are processed by the following stochastic computing logics which are determined by the given application. There are two major concerns to realize Bayesian inference applications on such system. One concern is that it usually requires large amount of SBGs because each evidence is represented by one SBG. As we have observed from many applications, especially with large scale, there are many evidences who have the same probabilities and may share the same SBGs to reduce the required SBG array scale. And the second concern is that it usually requires digital-to-analog converters (DACs) to convert the input digital sources into analog format which are defined as bias voltages~\cite{jia2017spintronics}. Meanwhile, the bias voltages margin is usually very small and high accuracy DACs are required to improve the input margin so that the design overheads are difficult to tolerant. In this work, SPINBIS is proposed to overcome these two disadvantages.

\begin{figure}[tb!]
    \centering
    \includegraphics[width=0.43\textwidth]{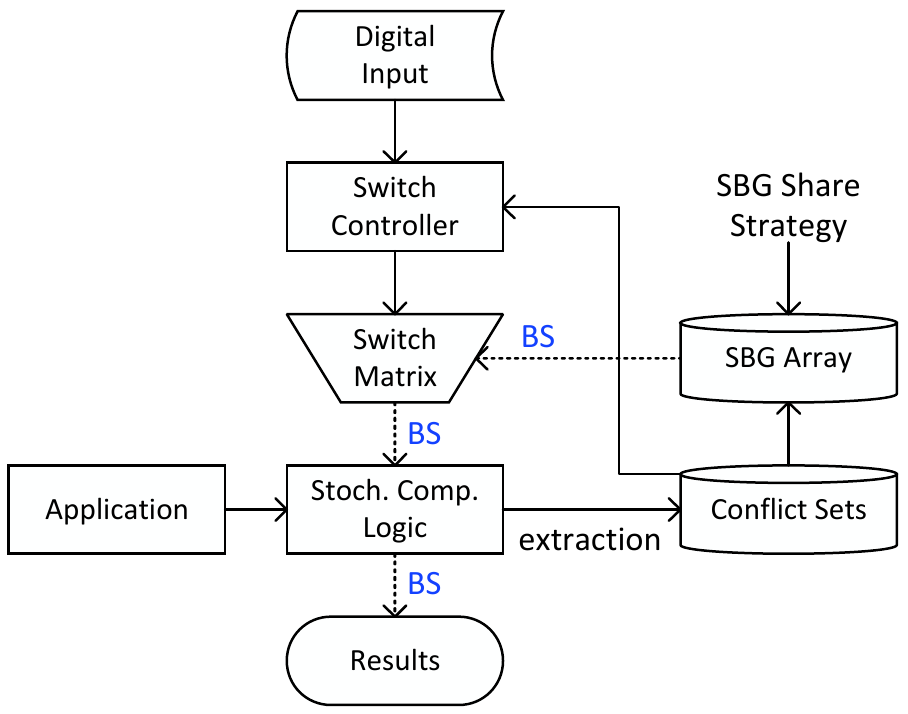}
    \caption{The proposed SPINBIS architecture.}
    \label{fig:spinbis}
\end{figure}

\subsection{Overview of SPINBIS} \label{Section:spinbis:overview}

SPINBIS is a spin-based Bayesian inference system. As shown in the diagram from Fig.~\ref{fig:spinbis}, an SBG sharing strategy is exploited in SPINBIS to significantly reduce the required array size which is different from the previous approach~\cite{jia2017spintronics}. The SBG sharing strategy allows the inputs with the same evidence could be potentially represented by the bitstream generated from the same SBG. However, there are some inputs who are connected together by one or more logic gates, which are regarded as \textit{conflicting} with each other. Conflicting inputs are not allowed to share the same SBG. As shown in Fig.~\ref{fig:spinbis}, the conflict sets are extracted from the stochastic computing logic which contains the conflicting relationship. The stochastic computing logic block is determined according to the specified applications. The SBG array is pre-built according to the specified applications and the generated bitstreams are assigned to stochastic computing logics by a switch matrix which is controlled by the digital inputs~\cite{masson1979sampler}. The input of switch matrix is the generated bitstreams from SBG array, and the output is connected to the stochastic computing logics. The switch matrix is a crossbar structure while each cross point is realized by a transistor which is controlled by the switch controller. In summary, the bitstreams from SBG array are assigned to stochastic computing logics according to the switch matrix which is controlled by switch controller.
% so that the total required number of SBGs is significantly reduced compared with Fig.~\ref{fig:aspdacarchi}.

\subsection{Stochastic Computing Logic and Conflict Set}\label{Section:spinbis:scl}

\begin{figure}[tb!]
    \centering
    \includegraphics{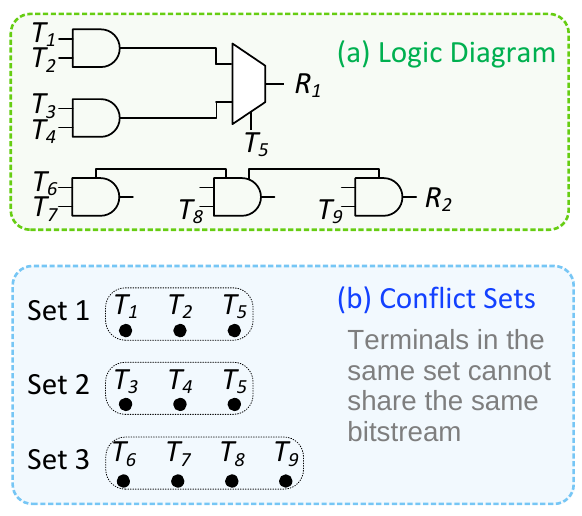}
    \caption{Stochastic computing logic diagram and its conflict set. Terminals of $T_1 \sim T_9$ are supposed to have the probabilities \{$p_1, p_2, p_1, p_3, p_1, p_4, p_5, p_3, p_3$\}.}
    \label{fig:scl_cg}
\end{figure}

One of the most attractive advantages of stochastic computing is that the involved arithmetic operations could be efficiently realized by simple logic gates, including \texttt{AND, MUX}, etc. The stochastic computing logics are determined according to the specified applications. Once the application is given, the stochastic computing logic is determined. For the determined computing logics with $N$ inputs, it requires $N$ bitstreams from the switch matrix.
% The proposed SBG sharing strategy allows part of the $N$ bitstreams (the inputs with the same evidence value) may accept the same bitstream source from SBG array.
As shown in Fig.~\ref{fig:scl_cg}(a) of a stochastic computing logics example, there are two independent sub-circuits with $9$ inputs ($T_1, T_2, \cdots, T_9$) and 2 outputs of $R_1$ and $R_2$. For a naive Bayesian inference system~\cite{jia2017spintronics}, it requires $9$ bitstreams from SBG array with $9$ SBG circuits.
% By adopting the SBG sharing strategy, the required number of SBG circuits could be reduced.

Suppose that $bs_i$ means the input bitstream of terminal $T_i$, and $p_i$ is the input probability of terminal $T_i$. $p(bs)$ means the corresponding probability of bitstream $bs$. The stochastic computing logics in Fig.~\ref{fig:scl_cg}(a) are built to realize~\eqref{eqn:cg_example}。
\begin{equation}
\label{eqn:cg_example}
\begin{aligned}
 p_{R1} &= p_{1} \cdot p_{2} \cdot p_{5} + p_{3} \cdot p_{4} \cdot (1 - p_{5})       \\
          &= p(bs_{1} \& bs_{2} \& bs_{5}) + p(bs_{3} \& bs_{4} \&  \overline{bs_{5}}) \\
 p_{R2} &= p_6 \cdot p_7 \cdot p_8 \cdot p_9 \\
          &= p(bs_6 \& bs_7 \& bs_8 \& bs_9)
\end{aligned}
\end{equation} As we have seen from Eqn.~\ref{eqn:cg_example}, \texttt{AND} operations are executed among $\{bs_1, bs_2, bs_5\}$ so that they are defined as conflicting with each other according to the stochastic computing principle. And $T_1, T_2, T_5$ are formulated as a conflict set as shown in Fig.~\ref{fig:scl_cg}(b). Similarly, $T_3, T_4, T_5$ are formulated as another conflict set as well as $T_6, T_7, T_8, T_9$. The input terminals in the same conflict set are not allowed to share the same bitstream source from SBG array even if they have the same input evidence. Otherwise, the input terminals with same input evidence are allowed to share the same bitstream.

\subsection{SBG Array and SBG Sharing Strategy}\label{Section:spinbis:sbgarray}

As shown in Fig.~\ref{fig:stochswitch}, MTJ switching probability is associated with bias voltage and duration time. Generally, either bias voltage or duration time is fixed and the other one is varied for random switching. In the previous approach~\cite{jia2017spintronics}, bitstreams are directly fed into stochastic computing logic from SBG array so that it usually requires many SBGs, as well as DACs. Furthermore, the output probability of SBG is highly sensitive to the input bias voltage whose margin is very small as reported in~\cite{jia2017spintronics}. Accurate mapping from digital probabilities to voltages requires DACs with high precision, and it is difficult to tackle the non-linear relationship between probabilities and voltages. More importantly, a slight noise or process variation may map a probability to an unexpected voltage. Aiming to overcome these limitations, the bitstreams in SPINBIS are provided with a pre-built SBG array and assigned to stochastic computing logic through a switch matrix.

Different with the SBG array in~\cite{jia2017spintronics}, a pre-build SBG array based on SBG sharing strategy is utilized in SPINBIS to improve the stability of SBG and reduce the required number of SBGs. The BL/SL of each SBG in the array is supplied by an internal voltage source that could provide more stable bias voltage than DACs. By this manner, the generated probability of each SBG is pre-determined and will be multiplexed by the switch matrix. According to the SBG sharing strategy, the required number of SBGs could be much smaller compared with the input terminals of stochastic computing logics because the non-conflicting terminals are allowed to share the same bitstreams. Since the SBG array is pre-built, it has to provide enough kind of bitstreams to satisfy the required accuracy of the stochastic computing which will be discussed later.

Assuming that it requires $L$ kinds of probabilities for a specified application, we define ${p_1, p_2, \cdots, p_L}$ as the required probabilities. Each kind of probability correspond to one SBG set which is denoted as $SBG_{i, \phi(i)}$, where $i=1, 2, \cdots, L$ is the index of each kind of probability set, and $\phi(i)$ is the required number of SBGs in each SBG set. For each SBG set $SBG_{i, \phi(i)}$, they generates the same probability $p_i$ but the bitstreams are different from each other. Let $M = \phi(1) + \phi(2) + \cdots + \phi(L)$, and $M$ denotes the total number of SBGs in SBG array. The SBG array is constructed based on the conflict sets and input probabilities. The conflict sets are pre-extracted from the stochastic computing logics according to the specified application. For a particular application, input probabilities could be evaluated and usually have a certain distribution which is adopted to determine the probability set in combination with the pre-extracted conflict sets.

Taking the example of stochastic computing logics in Fig.~\ref{fig:scl_cg},  input terminals of $T_1 \sim T_9$ are supposed to have the probabilities \{$p_1, p_2, p_1, p_3, p_1, p_4, p_5, p_3, p_3$\}, where $T_1$, $T_3$ and $T_5$ have the same probability $p_1$, $T_4$, $T_8$ and $T_9$ have the same probability $p_3$. Since $T_1$ and $T_3$ don't belong to the same conflict set, they could share the same bitstream from SBG array. But $T_5$ has to adopt the bitstream from other different SBG because it is conflicted with $T_1$ and $T_3$. Similarly, $T_4$ and $T_8$ could share the same bitstream but not for $T_9$. In this case, only $7$ SBGs are required in SPINBIS while $9$ SBGs are required if no SBG sharing strategy is utilized~\cite{jia2017spintronics}. Hence, the SBG sharing strategy could significantly reduce the required SBGs scale, especially for the applications with large scale of input probabilities.

\subsection{Switch Matrix and Switch Controller}\label{Section:spinbis:switchmatrix}

\begin{figure}[tb!]
    \centering
    \includegraphics{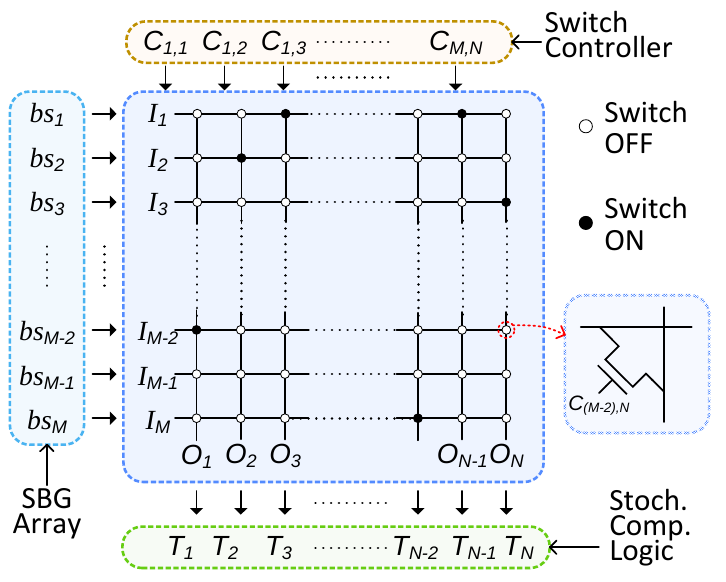}
    \caption{Switch matrix block of SPINBIS.}
    \label{fig:switchmatrix}
\end{figure}

The SBG sharing strategy is realized by exploiting a multiplexing network between SBG array and stochastic computing logics. As shown in Fig.~\ref{fig:switchmatrix}, the switch matrix receives the bitstreams from SBG array and assigns them to stochastic computing logics. The assigning procedures are determined by the switch controller. There are $M$ bitstream sources $bs_j$ to be linked to switch matrix left side terminal $I_j$, where $j = 1, 2, \cdots, M$. And there are $N$ outputs ($O_1, O_2, \cdots, O_N$) from switch matrix to be linked to the input terminals $T_k$ of stochastic computing logics, where $k = 1, 2, \cdots, N$. The switch matrix is built with a crossbar structure while nMOS transistor is located at each cross-point as a selector. The selection operations of these transistors are carried out by the switch controller which is determined by the digital inputs and conflict sets. For each column of the switch matrix, there is only one selector is switched ON because each input terminal of stochastic computing logics only accepts one bitstream. For each row of switch matrix, there may be zero, one or more selectors are switched ON because the bitstreams from SBG array may be shared by different input terminals of stochastic computing logics.

The switching procedures are illustrated in Alg.~\ref{alg:controlsignal}. In Lines (\ref{alg:controlsignal:init_begin}-\ref{alg:controlsignal:init_end}), the vector $bs[i]$ indicates the first available bitstream index of probability $p_i$. Lines (\ref{alg:controlsignal:generation_begin}-\ref{alg:controlsignal:generation_end}) generate control signals for all terminals in the given conflict set. For the terminals in one conflict set, the digital inputs (Line~\ref{alg:controlsignal:pro}) are obtained by the terminal index (Line~\ref{alg:controlsignal:terminal_index}). Then the probability index in vector $P$ is calculated by Line~\ref{alg:controlsignal:pro_index}. The control signal is generated by Line~\ref{alg:controlsignal:generation}. In Line~\ref{alg:controlsignal:updata_bs}, the first available bitstream index of the probability is updated. By this way, it could guarantee that each bitstream will not be allocated for terminals who belong to the same conflict set.

\begin{algorithm}[tb!]
% \small
\caption{Switching procedures for SBG sharing strategy.}
\label{alg:controlsignal}
\begin{algorithmic}[1]
    \Require
        Digital inputs $In[i]$, where $i=1,2, \cdots, N$;
        SBG array $SBG_{i,\phi(i)}$ with $P[i] = p_i$, and $\phi[i] = \phi(i)$, where $i=1,2, \cdots, L$;
        Conflict sets $cflct[i]$, where $i=1,2, \cdots, T$.
    \Ensure
        Binary control signal $C[i][j]$, where $i=1,2, \cdots, M$ and $j = 1,2, \cdots, N$.
    \State $bs[1] = 1$  \Comment $bs[i]$ indicates the first available bitstream index of probability $p_i$ \label{alg:controlsignal:init_begin}
    \For {($i = 2, i \le L; i = i + 1$)}
        \State $bs[i] = bs[i-1] + \phi[i-1]$
    \EndFor                                                                                                 \label{alg:controlsignal:init_end}
    \For {($i = 1, i \le T; i = i + 1$)}                                                                    \label{alg:controlsignal:generation_begin}
        \State $ter\_idx = cflct[i]$                                                                        \label{alg:controlsignal:terminal_index}
        \State $pro = In[ter\_idx]$                                                                         \label{alg:controlsignal:pro}
        \State $pro\_idx = findProIndex(pro, P)$                                                            \label{alg:controlsignal:pro_index}
        \State $C[bs[pro\_idx]][ter\_idx] = 1$                                                              \label{alg:controlsignal:generation}
        \State $bs[pro\_idx] = bs[pro\_idx] + 1$                                                            \label{alg:controlsignal:updata_bs}
    \EndFor                                                                                                 \label{alg:controlsignal:generation_end}
\end{algorithmic}
\end{algorithm}

Even though SBG sharing strategy has been utilized to reduce the required scale of SBG array, the scale of switch matrix is still too large because the stochastic computing logics usually have too many input terminals. In this work, a terminal clustering strategy is further proposed to reduce the scale of switch matrix. For the input terminals of stochastic computing logics who \textit{always} have the same digital input, they are clustered as a single terminal if they are in the different conflict sets. As shown in Fig.~\ref{fig:scl_cg}, terminals $T_1$ and $T_3$ belong to different conflict sets, if they always have the same input probability, they are clustered as the same input terminal.

\subsection{Discussion}\label{Section:spinbis:discussion}

The switch matrix and SBG sharing strategy is proposed in SPINBIS to reduce the required number of SBGs with certain design overhead. We compare the design complexity between SPINBIS and the work in~\cite{jia2017spintronics}. The stochastic computing logics of SPINBIS are the same as~\cite{jia2017spintronics}. The scale of SBG array is reduced from $N$ to $M$ according to the SBG sharing strategy, where $M \ll N$. Since the SBG array accounts for substantial part of energy consumption in SPINBIS, the energy consumption is reduced by $\frac {N-M} {N}$ when the scale of SBG array is reduced from $N$ to $M$. Assuming that there are $T$ transistors in each SBG circuit, the utilization of transistors in SBG array is reduced from $T * N$ to $T * M$. According to the terminal clustering strategy, the number of switch matrix output $N$ is reduced as $N'={\alpha}N$, where $\alpha \in (0,1)$, and the utilization of transistors in switch matrix is $M * {\alpha}N$. In summary, the utilization of transistors of SBG array is reduced from $T * N$ to $T * M$ but with the overhead of $M * {\alpha}N$ resulted from switch matrix. Since $M \ll N$, the total area of SPINBIS $\left( T * M + M * {\alpha}N \right)$ is mainly determined by $M * {\alpha}N$. Based on the above discussion, the advantages of SPINBIS can be well highlighted when dealing with large scale applications with regular structure and input patterns.

% {\color{blue}Based on the above analysis, the advantages of SPINBIS can be better highlighted when dealing with applications with regular structure and input. If it is used to solve Bayesian belief network or Bayesian neural network which contains many well-trained parameters, SPINBIS does not have much advantage, and even increases the power consumption and area of the system. In order to make SPINBIS more generally applied, architectures illustrated in Fig.~\ref{fig:aspdacarchi} and Fig.~\ref{fig:spinbis} could be combined as a hybrid one. In the hybrid architecture, there are two class of SBG arrays. The first class is the same as it in SPINBIS. For those stochastic computing logic terminals whose input value vary for different cases, they are connected with the first class SBG array by switch matrix. The second class SBG array is similar to that in~\cite{jia2017spintronics}. For those stochastic computing logic terminals whose value is constant (such as well-trained weights in neural network), they are directly connected with the second class SBG array. By this method, the voltage setting of the second class SBG array is fixed. It does not require DACs and will not be seriously affected by noise and process variation. }

 \iffalse

 \bibliography{../ref/Bayesian}

 \fi

\section{Spintronic Device based Energy Efficient SBG}\label{Section:sbg}

The performance of SBG is critical for efficient SPINBIS both in inference accuracy and inference speed as well as the power consumption. A high quality SBG should have the following two properties at least: (1) The generated bitstream could represent the given probability as accurately as possible. If the deviation between probability value and bitstream is too large, the stochastic computing results will be unpredictable. (2) The correlations among different stochastic bitstreams should be as small as possible because high correlation usually degrade the accuracy of stochastic computing significantly. In this section, an efficient SBG circuit is proposed by utilizing the inherent random behaviors of MTJ devices for Bayesian inference.

\subsection{Schematic of SBG}\label{Section:sbg:General}

% \begin{figure}[tb!]
%     \centering
%     \includegraphics[scale=1]{statetransitionall.eps}
%     \caption{State transition diagram of (a) simple SBG and (b) self-control SBG.}
%     \label{fig:statetransitionGeneral}
% \end{figure}

\begin{figure}[tb!]
    \centering
    \includegraphics[scale=0.9]{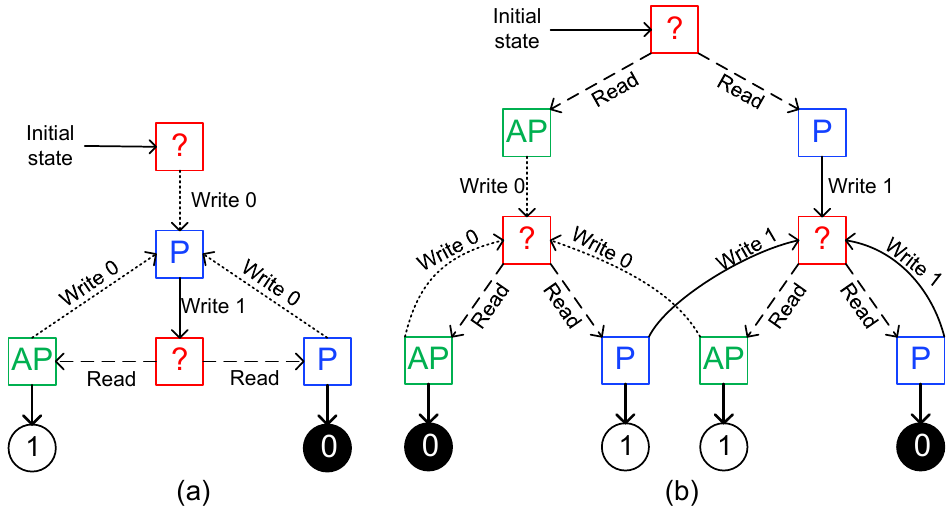}
    \caption{State transition diagram of (a) simple SBG and (b) self-control SBG.}
    \label{fig:statetransition}
\end{figure}

\begin{figure*}[tb!]
    \centering
    \includegraphics[width=0.9\textwidth]{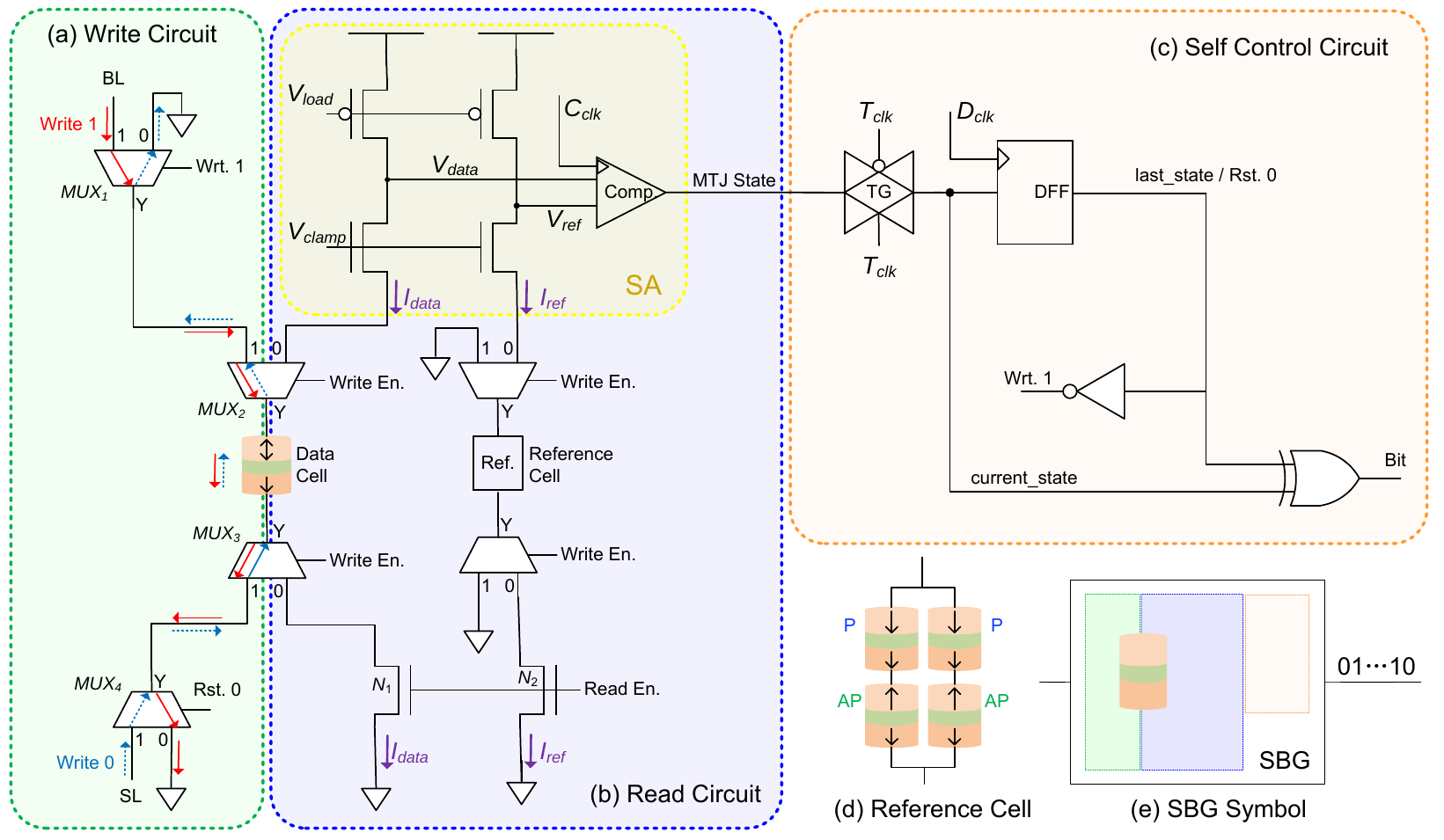}
    \caption{Schematic of SBG circuit.}
    \label{fig:sbg_schematic}
\end{figure*}

The stochastic bitstreams are generated by reading the MTJ states which have been pre-written as shown in Fig.~\ref{fig:stochswitch}. If the readout of MTJ is with high resistance \textit{i.e.} `\texttt{AP}' state, `1' will be generated as one stochastic bit; otherwise, `0' will be generated. Generally, each bit generation is accomplished by three stages: \textit{reset}, \textit{write} and \textit{read}. Bitstreams are obtained by performing these three stages continuously. The state transition diagram of simple SBG is illustrated in Fig.~\ref{fig:statetransition}(a).

Both the \textit{reset} procedure and \textit{write} procedure is a kind of programming operation on MTJ device. The \textit{reset} operation aims to program the MTJ with bias voltage and duration time which is large enough to achieve a successful switching while the switching probability is close to $100\%$. But the \textit{write} operation aims to program the MTJ according to the required switching probability ($p \in \left[ {0,1} \right]$) as shown in Fig.~\ref{fig:stochswitch}. Assuming that the initial MTJ state is unknown, the \textit{reset} operation (\textit{Write 0} in Fig~\ref{fig:statetransition}(a)) is to switch it to `\texttt{P}' state with the probability $p = 100\%$ while the \textit{write} operation (\textit{Write `1'} in Fig~\ref{fig:statetransition}(a)) is to switch it with the probability $p \in \left[ {0,1} \right]$.

\begin{table}[tp]
\centering
\caption{Enable signal configuration for \textit{reset}, \textit{write} and \textit{read} operations.}
\label{Table:operation}
\begin{threeparttable}
\begin{tabular}{p{1.2cm}<{\centering}!{\VRule[0.8pt]}p{1.2cm}<{\centering}!{\VRule[0.8pt]}p{1.2cm}<{\centering}!{\VRule[0.8pt]}p{1.2cm}<{\centering}!{\VRule[0.8pt]}p{1.2cm}<{\centering}}
    \specialrule{0.8pt}{0pt}{0pt}
                    & \textit{Write En.}       & \textit{Read En.}         & \textit{Rst. 0}  &  \textit{Wrt. 1} \\
    \hline
    \textit{reset}  &  \multirow{2}{*}{High}   &   \multirow{2}{*}{Low}    &     High         &    Low      \\
    \cline{1-1}\cline{4-5}
    \textit{write}  &                          &                           &     Low          &    High     \\
    \hline
    \textit{read}   &  Low                     &   High                    &     -            &    -        \\
    \specialrule{0.8pt}{0pt}{0pt}
\end{tabular}
\end{threeparttable}
\end{table}

The enable signal configuration has been illustrated in Table~\ref{Table:operation}. Both the \textit{write} and \textit{reset} operations are accomplished by the write circuit as shown in Fig.~\ref{fig:sbg_schematic}(a) while the \textit{read} operation is finished by the read circuit as shown in Fig.~\ref{fig:sbg_schematic}(b). The multiplexers \texttt{MUX}$_2$ and \texttt{MUX}$_3$ are adopted to switch the write current or read current flowing through the MTJ.

During \textit{write} and \textit{reset} operations, \textit{Write En.} is set as high, thus terminal `1' of \texttt{MUX}$_2$ and \texttt{MUX}$_3$ are connected with corresponding terminal `Y'. For \textit{reset} operation, \textit{Wrt. 1} is set as low and \textit{Rst. 0} is set as high so that terminal `0' of \texttt{MUX}$_1$ and terminal `1' of \texttt{MUX}$_4$ are connected with terminal `Y'. By applying a bias voltage between source-line (SL) and \textit{GND}, write current flows through the MTJ from bottom to top as the blue arrow shows. For \textit{write} operation, \textit{Wrt. 1} is set as high and \textit{Rst. 0} is set as low so that terminal `1' of \texttt{MUX}$_1$ and terminal `0' of \texttt{MUX}$_4$ are connected with terminal `Y'. By applying a bias voltage between bit-line (BL) and \textit{GND}, current flows through the MTJ from top to bottom as the red arrow shows.

During \textit{read} stage, \textit{Read En.} is set as high while \textit{Write En.} is set as low so that terminal `0' of \texttt{MUX}$_2$ and \texttt{MUX}$_3$ are connected with terminal `Y'. A pre-charging sense amplifier is adopted to compare the MTJ state of data cell with that of the reference cell as shown in Fig.~\ref{fig:sbg_schematic}(b). The MTJ resistance state of reference cell (\ref{fig:sbg_schematic}(d)) is usually set as $(R_P+R_{AP})/2$ so that both \texttt{AP} state and \texttt{P} state of data cell could be identified correctly. The read circuit consists of a two-branch sensing circuit with equalizing transistors~\cite{kim2012novel} and a voltage sense amplifier with dynamic latched comparator~\cite{figueiredo2009offset} for digital output. Both branches of read circuit are composed by a load pMOS, a read enable nMOS and a clamped nMOS~\cite{yang2016radiation}\cite{yang2019exploiting}. The \textit{read} operation is enabled by setting \textit{Read En.} as high so that nMOS $N_1$ and $N_2$ are turned on. The clamped nMOS is utilized to prevent read disturbance by applying a proper bias voltage $V_{clamp}$. The resistance of reference cell is usually located between $R_P$ and $R_{AP}$ in order to identify the `\texttt{AP}' state or `\texttt{P}' state of data cell. During \textit{read} stage, the resistance difference between date cell and reference cell is converted to the difference of $V_{data}$ and $V_{ref}$ which could be sensed by a dynamic latched voltage comparator with clock enabled. The state of data cell is read out at each rising edge of $C_{clk}$.

% \begin{figure}[tb!]
%     \centering
%     \includegraphics[scale=1]{statetransition2.eps}
%     \caption{State transition diagram of self-control SBG.}
%     \label{fig:statetransitionImproved}
% \end{figure}

\subsection{Energy Efficient SBG Using Self-control Strategy}\label{Section:sbg:Improved}

Energy efficiency of neural network and Bayesian inference has been considered as a primary concern for applications on embedded computing platforms. Several research works have been proposed towards efficient implementation of MTJ-based stochastic computing. The work in \cite{mondal2017power} indicated that the energy consumption required for switching \texttt{P}$\rightarrow$\texttt{AP} with $99.9\%$ probability is less than that of switching \texttt{AP}$\rightarrow$\texttt{P}. Hence, they reset the MTJ to \texttt{AP} state every time and then attempt to switch it to \texttt{P} state to generate one stochastic bit. However, the energy consumption of resetting \texttt{P}$\to$\texttt{AP} is still wasted because no bit is generated during the \textit{reset} procedure. As illustrated in Fig.~\ref{fig:statetransition}(a) of simple SBG, the MTJ is first reset as `\texttt{P}' state after each stochastic bit is generated. The stochastic bits are generated by reading out the MTJ state after the \textit{write} procedure. Actually, the stochastic bits are generated based on whether the MTJ state is switched successfully or not in \textit{write} procedure. In this work, we propose an efficient SBG while the \textit{reset} procedure is also utilized for generating stochastic bits.

A self-control strategy is proposed by storing the MTJ state of previous cycle in a register and then comparing it with the state of current cycle to determine the stochastic bit as output. That is, the stochastic bit is generated according to the comparison whether MTJ state is changed or not. The state transition diagram of SBG with self-control strategy is illustrated in Fig.~\ref{fig:statetransition}(b). If the current state is different from the last state, the output bit is `1', otherwise, the output bit is `0'. Meanwhile, the direction of \textit{write} operation is determined by the stored state of last cycle. According to the self-control strategy, the \textit{reset$\to$write$\to$read} procedures are compressed as \textit{write$\to$read} procedures. In \textit{write} procedure, the biased voltage between BL and SL are carefully set as a certain range to guarantee both \textit{write `0'} and \textit{write `1'} operations are with the same probability value. The speed and energy efficiency of bitstream generation could be improved by $2\times$ theoretically.

The self-control circuit is demonstrated in Fig.~\ref{fig:sbg_schematic}(c). The transmission gate (TG) and D-Flip-Flop (DFF) is clocked by $T_{clk}$ and $D_{clk}$, respectively. The output of comparator in Fig.~\ref{fig:sbg_schematic}(b) (i.e. MTJ State) is highly sensitive to its loads. Hence, the transmission gate is inserted to eliminate the loads influence. There is a small delay in rising edge of $T_{clk}$ after the rising edge of $C_{clk}$ to guarantee the output of comparator is stable. DFF is utilized to latch the MTJ state which will be compared with next cycle for one stochastic bit output. If the current MTJ state is different from the latched state, the output of SBG is `1', otherwise, is `0'. Meanwhile, the latched MTJ state also determines the direction of \textit{write} operation in the next cycle. If the latched MTJ state is \texttt{P}, the write current flows through the MTJ from top to bottom which attempts to switch the MTJ state from \texttt{P} to \texttt{AP}. Otherwise, the \textit{write} current has the opposite direction. There is also a small delay in rising edge of $D_{clk}$ after the rising edge of $T_{clk}$. When $T_{clk}$ is high and $D_{clk}$ is low, TG output is the current state of MTJ and DFF output is the last state of MTJ. During this period, XOR operation on current state and last state is regarded as one bit output. If the current state is different from the latched state in the last cycle, it means that the state of MTJ has already switched successfully. The result of the XOR gate is high, and consequently one bit of `1' is generated. Otherwise, one bit `0' is generated. After the rising edge of $D_{clk}$, current state is latched in DFF until the next \textit{read} stage. In next \textit{write} stage, the DFF output \textit{Rst. 0} is utilized to control \texttt{MUX}$_4$ and \textit{Wrt. 1} is utilized to control \texttt{MUX}$_1$.

\begin{figure}[tb!]
    \centering
    \subfloat[] {
    \includegraphics[scale=1.4]{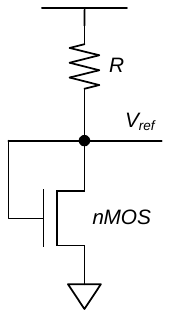}
    \label{fig:voltagedivider:a}
    }
    \subfloat[] {
    \includegraphics[scale=0.38]{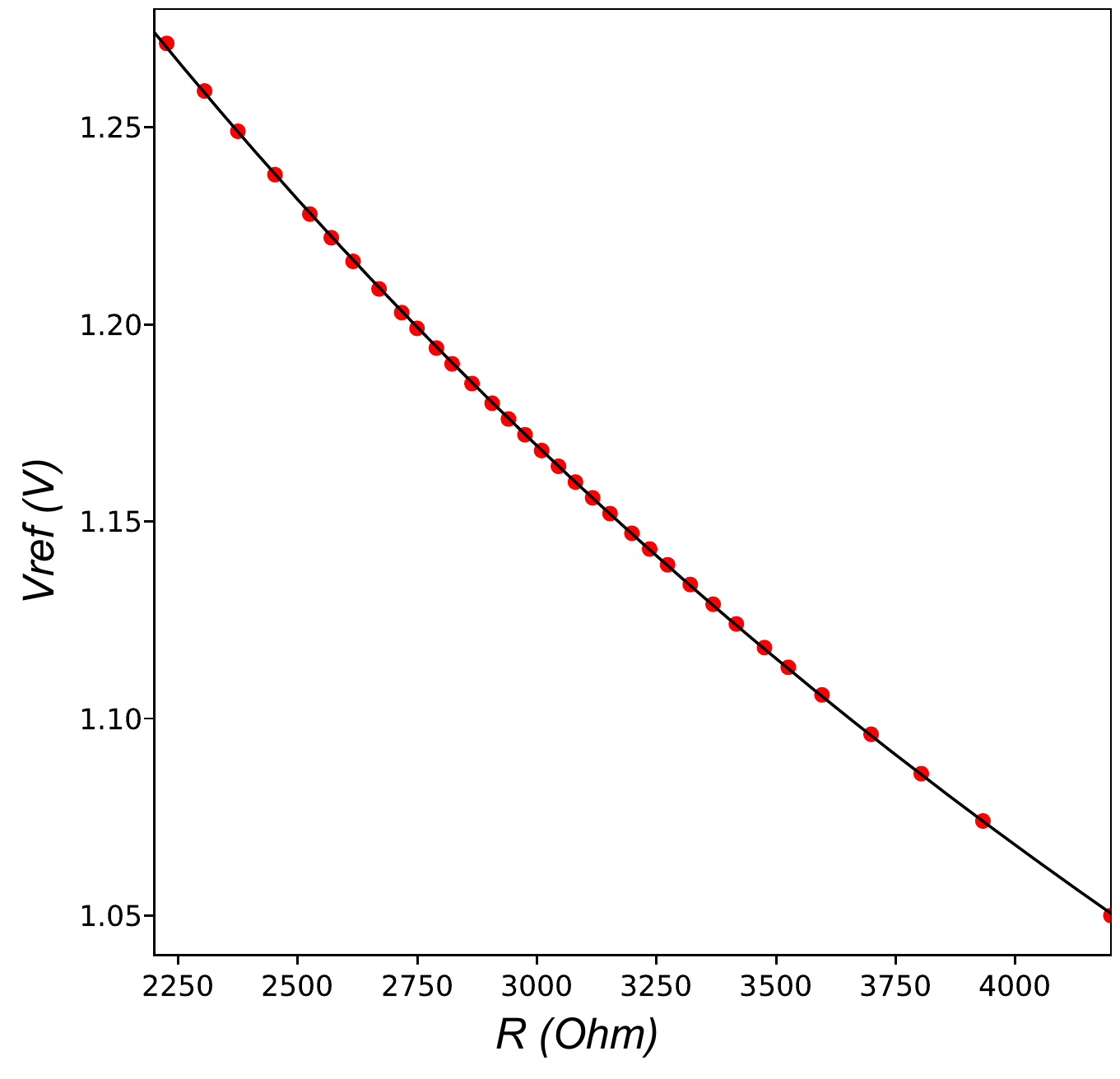}
    \label{fig:voltagedivider:b}
    }
    \caption{a) Schematic of voltage divider. b) The black solid line represents the fitting curve of $R$ and $V_{ref}$, the red circles represent the desired writing voltages between BL and SL of SBGs.}
    \label{fig:voltagedivider}
\end{figure}

The aforementioned SPINBIS architecture requires that each SBG should be equipped with two internal voltage sources with fixed voltage values. In this work, it is achieved by a voltage divider~\cite{baker2008cmos} that consists of one resistor and one nMOS transistor as shown in Fig.~\subref*{fig:voltagedivider:a}. $V_{ref}$ is adopted as the writing voltage for each SBG, and determined by varying the resistance value $R$. As shown in Fig.~\subref*{fig:voltagedivider:b}, $V_{ref}$ varies smoothly according to $R$ (black solid line). And the desired writing voltage between BL or SL (red circles in Fig.~\subref*{fig:voltagedivider:b}) could be achieved by adjusting the resistor value.

For the sake of convenient, the SBG with self-control strategy is denoted as \texttt{self-control SBG} and the one described in Section~\ref{Section:sbg:General} is denoted as \texttt{simple SBG}.

\subsection{Evaluation of SBG Circuits}\label{Section:sbg:evaluation}

The proposed SBG circuits are evaluated to explore its performance in this section.

\subsubsection{Simulation setup}

The SBG circuit is composed by hybrid CMOS/MTJ structures with $45 \; nm$ CMOS and $45 \; nm$ MTJ technologies. A behavioral model of MTJ is described by Verilog-A language~\cite{wang2014compact} while the stochastic switching behaviors are also included. However, the original MTJ model in~\cite{wang2014compact} only provides stochastic switching behaviors for a single device. That is, the obtained bitstreams from different SBG circuits are always the same if they have a same bias voltage and duration time. Since many MTJs are utilized in SBG array, the bitstreams generated with original MTJ model ~\cite{wang2014compact} will have very strong correlation with each other which will lead to inaccuracy stochastic computing results. Hence, a new compact MTJ model is proposed for stochastic switching with the property that the switching behaviors of different MTJ instances are different with each other.

\begin{table}[tb]
\centering
\caption{The parameters definition and default value of MTJ model.}
\label{Table:MTJParam}
\begin{tabular}{l|l|l}
    \specialrule{0.8pt}{0pt}{0pt}
     Parameter & Description & Default value \\
    \hline
     $\alpha$ & Gilbert damping coefficient &      $0.027$ \\
     $\gamma$ & Gyro-Magnetic constant & $1.76 \times {10^7} \; Hz/Oe$ \\
     $P$ & Electron polarization percentage &       $0.52$ \\
     $H_{k0}$ & Out of plane magnetic anisotropy &    $1433 \; Oe$ \\
     $t_{sl} / t_{ox}$ & Height of the free layer / oxide barrier   &  $1.3 \; nm$ / $0.85 \; nm$ \\
     $l / w$ & Length / width of MTJ traverse &   $45 \; nm$ / $45 \; nm$ \\
     % $w$ & Width of MTJ traverse &   $45 \; nm$ \\
     % $t_{ox}$ & Height of the oxide barrier & $0.85 \; nm$ \\
     $TMR$ & TMR with zero volt bias voltage &        $1.5$ \\
     $RA$ & Resistance area product &     $5 \; \Omega \cdot μm^2$ \\
     \specialrule{0.8pt}{0pt}{0pt}
\end{tabular}
\end{table}

\begin{figure}[tb!]
    \centering
    \includegraphics[width=0.5\textwidth]{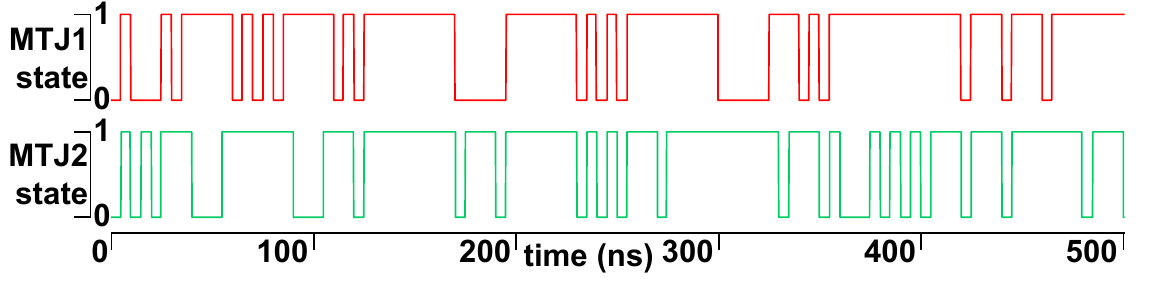}
    \caption{Simulation results of the proposed instance-vary MTJ model. Two MTJs are simulated simultaneously with the same bias voltage and pulse width.}
    \label{fig:randomseedresults}
\end{figure}

As described in~\cite{wang2014compact}, MTJ switching time is obtained by the critical current and other electrical and physical parameters. The stochastic behavior is independent with the critical switching current, and is implemented by random functions with uniform or normal distributions. The basic switching time $dt$ is determined according to the applied bias voltage for each cycle. Then a random number that obeys normal distribution $\sim N(seed, dt, \sigma)$ is generated per cycle as the final switching time, where $seed$ is the random seed for random number generation function, $\sigma$ is the user specified standard deviation. The parameter $seed$ is set as a constant value in the model published in~\cite{wang2014compact}. It indicates that the switching time is the same if two MTJs have the same $dt$ and $\sigma$. Aiming to obtain the different random behaviors, the MTJ model is revised by setting $seed$ as different values for different MTJ instances, which is denoted as \textit{instance-vary} model. Fortunately, the Verilog-A language of version 13.1 and above supports the grammar of \texttt{arandom[param]}. The \texttt{param} argument is optional and can be set as \texttt{global} or \texttt{instance}. If \texttt{param} is set as \texttt{instance} for each MTJ's required randomness, different $seed$ values will be generated for different MTJ instances. This feature could satisfy the MTJ's \textit{instance-vary} randomness requirement well. The parameters definition and default value of MTJ model are provided in Table~\ref{Table:MTJParam} for experimental configurations. The simulation results of MTJ switching with \textit{instance-vary} model are shown in Fig.~\ref{fig:randomseedresults}. With the same \textit{write} operations, MTJ1 and MTJ2 have different switching results which are critical to generate irrelevant bitstreams for stochastic computing.

\subsubsection{SBG simulation results}

\begin{figure}[tb!]
    \centering
    \includegraphics[width=0.5\textwidth]{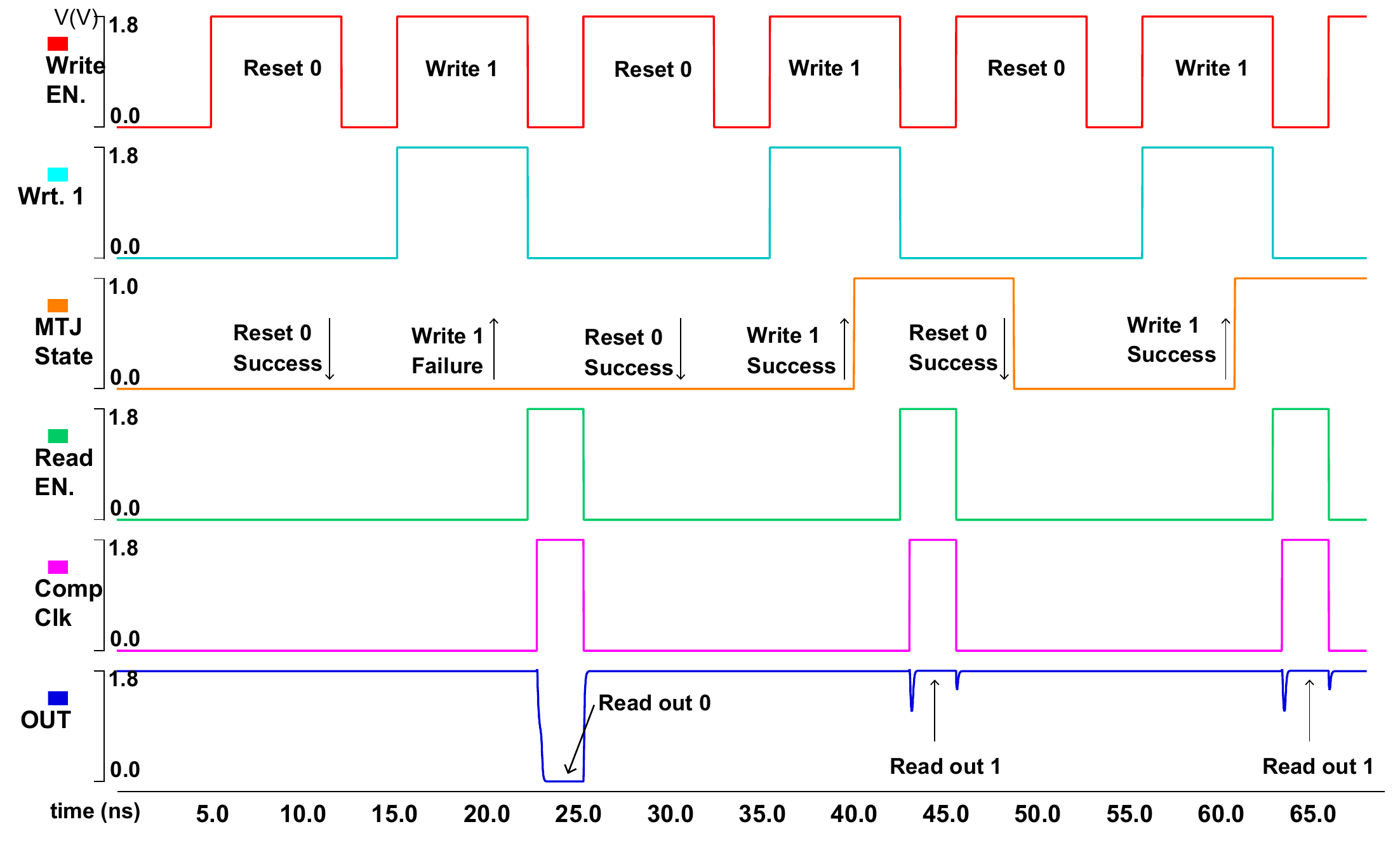}
    \caption{Simulation results of simple SBG circuit.}
    \label{fig:wave}
\end{figure}

The simulation results of simple SBG circuit are illustrated in Fig.~\ref{fig:wave}. The \textit{reset$\to$write$\to$read} operations are performed iteratively for 3 cycles from $5$ $ns$ to $65$ $ns$. The \textit{reset} and \textit{write} operations are enabled when \textit{Write En.} is high. For \textit{reset} operation (\texttt{AP} $\to$ \texttt{P} or \texttt{P} $\to$ \texttt{P}), the bias voltage between SL and \textit{GND} is about $1.8$ $V$ and the duration time is about $7$ $ns$ to guarantee the switching probability $p \to 100\%$. For \textit{write} operation (\texttt{P} $\to$ \texttt{AP}), if the bias voltage between BL and \textit{GND} is set as about $1.166$ $V$ and duration time is about $5.4$ $ns$, the switching probability $p$ is about $50\%$. For \textit{read} operation, \textit{Read En.} is set as high and the MTJ state is read out while $V_{load}$ and $V_{clamp}$ is about $0.8$ $V$. For each cycle of \textit{reset$\to$write$\to$read} operations, the MTJ is first reset as \texttt{P} state with the switching probability $p \to 100\%$ when \textit{Write En.} and \textit{Wrt. 1} is set as low. The write current flows through the MTJ from bottom to up as the blue arrow shown in Fig.~\ref{fig:sbg_schematic}(a). And then the MTJ attempts to finish \texttt{P} $\to$ \texttt{AP} switching with the provided switching probability when \textit{Wrt. 1} is set as high. The write current flows through the MTJ from up to bottom as the red arrow shows in Fig.~\ref{fig:sbg_schematic}(a). At last, \textit{read} operation is performed by setting \textit{Read En.} as high and \textit{Write En.} as low. For the 3 cycles of \textit{reset$\to$write$\to$read} operations as shown in Fig.~\ref{fig:wave}, writing \texttt{P} $\to$ \texttt{AP} fails in the first cycle but successes in the following two cycles. And consequently, the bitstream is generated as `011'.

\begin{figure}[tb!]
    \centering
    \includegraphics[width=0.47\textwidth]{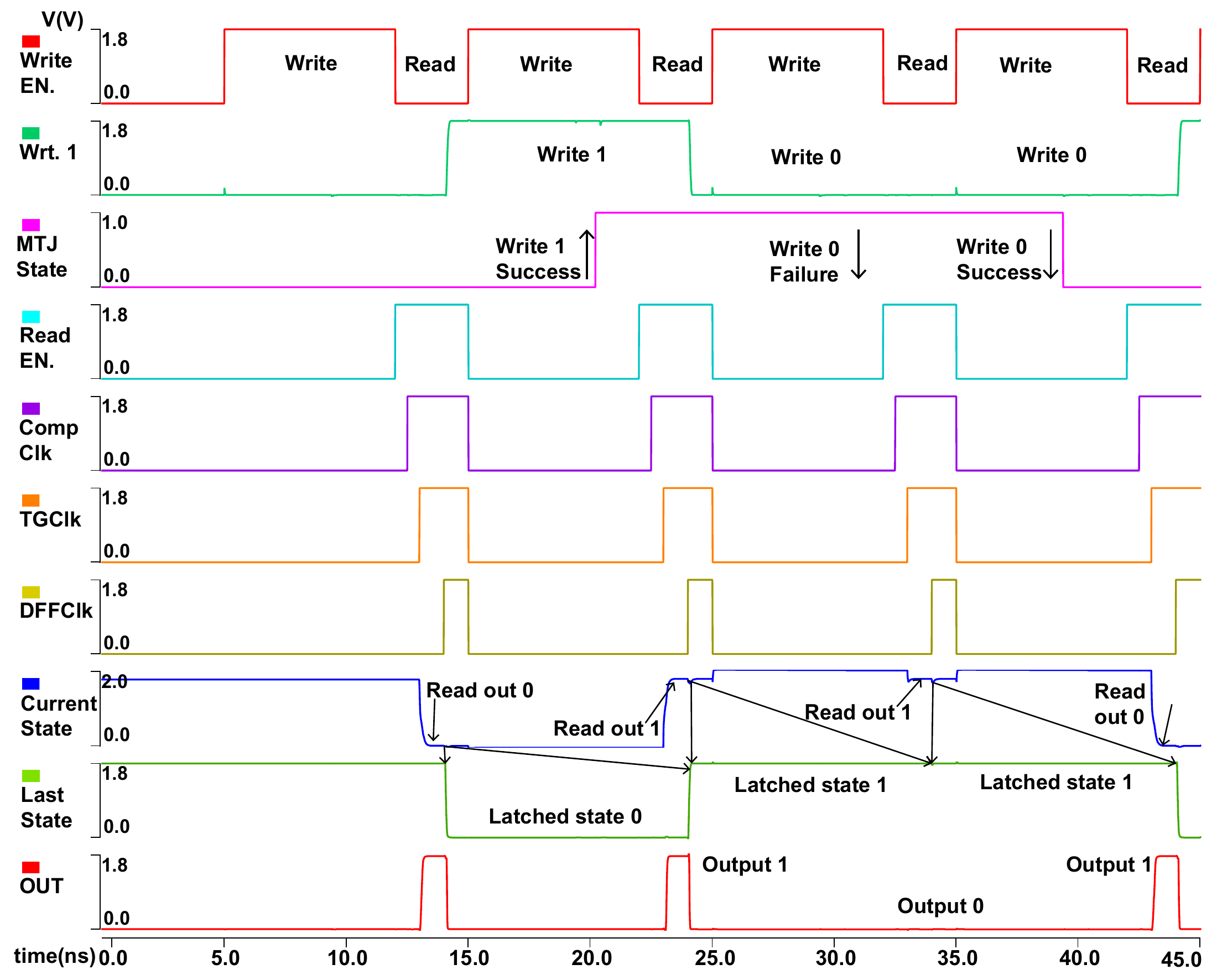}
    \caption{Simulation wave of self-control SBG circuit.}
    \label{fig:waveselfctrl}
\end{figure}

The comprehensive simulation results of self-control SBG circuit are shown in Fig.~\ref{fig:waveselfctrl} for 4 cycles from 5 $ns$ to 45 $ns$. The first cycle aims to initialize MTJ as \texttt{P} state. The MTJ is read out as \texttt{P} state at $13$ $ns$ and TG is turned on with a delay of $0.5$ $ns$. Since the `last\_state' is meaningless for the first cycle, the XOR result is discarded in this cycle. And then the current state \texttt{P} is latched in DFF from $14$ $ns$ to $24$ $ns$. For the second cycle, \textit{Wrt. 1} is enabled for a \textit{write} operation while `last\_state' is \texttt{P} (logic `0'). The \textit{write} operation is finished successfully so that the MTJ is in \texttt{AP} state. From $23$ $ns$ to $24$ $ns$, the \texttt{AP} state of MTJ is passed through TG and denoted as `current\_state'. By performing XOR operation on the `last\_state' (latched \texttt{P}) and `current\_state' (\texttt{AP}), one bit of `1' is generated for the second cycle. The `current\_state' (\texttt{AP}) is then latched in DFF and becomes as the `last\_state' for the next cycle. For the third cycle, \textit{Rst. 0} is enabled for writing MTJ from \texttt{AP} to \texttt{P} state since the latched `last\_state' is in \texttt{AP} state. However, the writing operation of \texttt{AP} to \texttt{P} fails for the third cycle. It means that the MTJ state in the third cycle is not changed compared with that in the second cycle. Hence, one bit of `0' is generated for the third cycle. For the forth cycle, MTJ still attempts to switch from \texttt{AP} to \texttt{P} state and finishes the switching successfully. Hence, one bit of `1' is generated. Finally, a bitstream `101' is obtained among these 4 cycles.

For generating a bitstream with $n$ bits, simple SBG circuit requires $2n$ \textit{write} operations (including \textit{reset} and \textit{write}) and $n$ \textit{read} operations. But for self-control SBG circuit, only $n+1$ \textit{write} operations (including initialization and \textit{write}) and $n+1$ \textit{read} operations are required. It is obvious that the self-control SBG circuit could improve the speed and energy efficiency about $2\times$ compared with simple SBG circuit.

\subsubsection{SBG performance}

The proposed SBG circuits are evaluated to analyze the performance of the generated stochastic bitstreams both on representation accuracy and correlation.

\begin{figure}[tb!]
    \centering
    \includegraphics[width=0.48\textwidth]{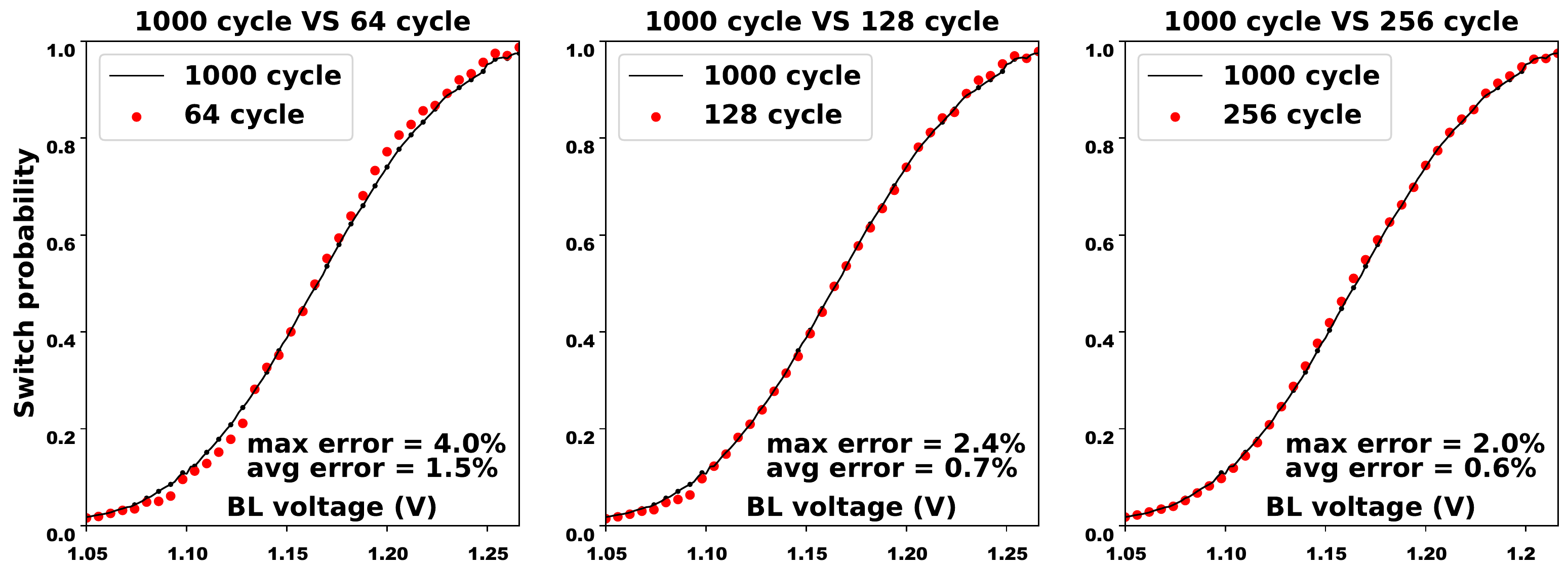}
    \caption{MTJ switching probability with different applied BL voltage. SL voltage is set as $1.8$ $V$.}
    \label{fig:pvrelation_bl}
\end{figure}

\begin{figure}[tb!]
    \centering
    \includegraphics[width=0.48\textwidth]{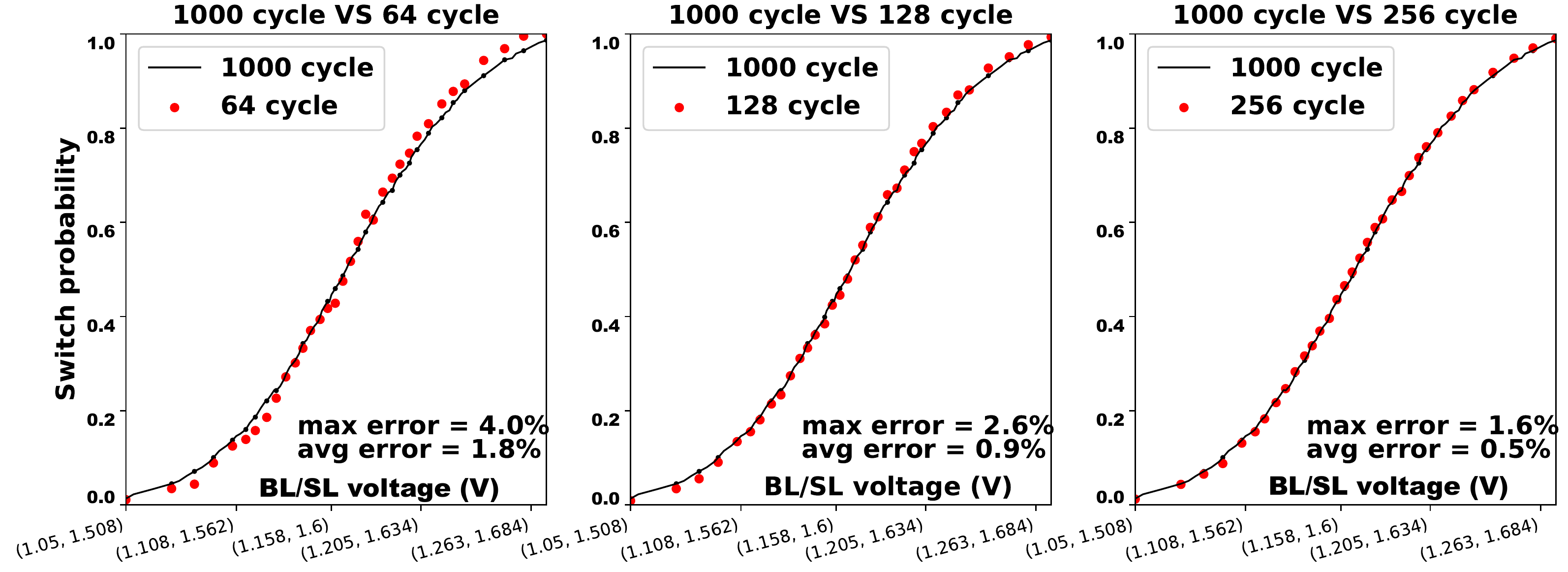}
    \caption{MTJ switching probability with different applied BL/SL voltage combination.}
    \label{fig:pvrelation_blsl}
\end{figure}

For evaluating the accuracy, $n$-bits stochastic bitstreams are generated while $n$ is the bitstream length of $64$, $128$, $256$ and $1000$. The bitstream with length $n = 1000$ is denoted as the ground truth. The MTJ switching probability of simple SBG is demonstrated in Fig.~\ref{fig:pvrelation_bl} with different BL voltage while the SL voltage is set as $1.8 \; V$. Compared with the ground truth of length $n = 1000$, the generated bitstreams with length $n = 64, 128, 256$ have the average errors of $1.5\%$, $0.7\%$ and $0.6\%$, respectively. Meanwhile, the relationship between switching probability and different BL/SL voltage combinations are also demonstrated in Fig.~\ref{fig:pvrelation_blsl} for self-control SBG circuit. Compared with the ground truth of length $n = 1000$, the generated bitstreams with length $n = 64, 128, 256$ have the average errors of $1.8\%$, $0.9\%$ and $0.5\%$, respectively.

\begin{figure}[tb!]
    \centering
    \includegraphics[width=0.38\textwidth]{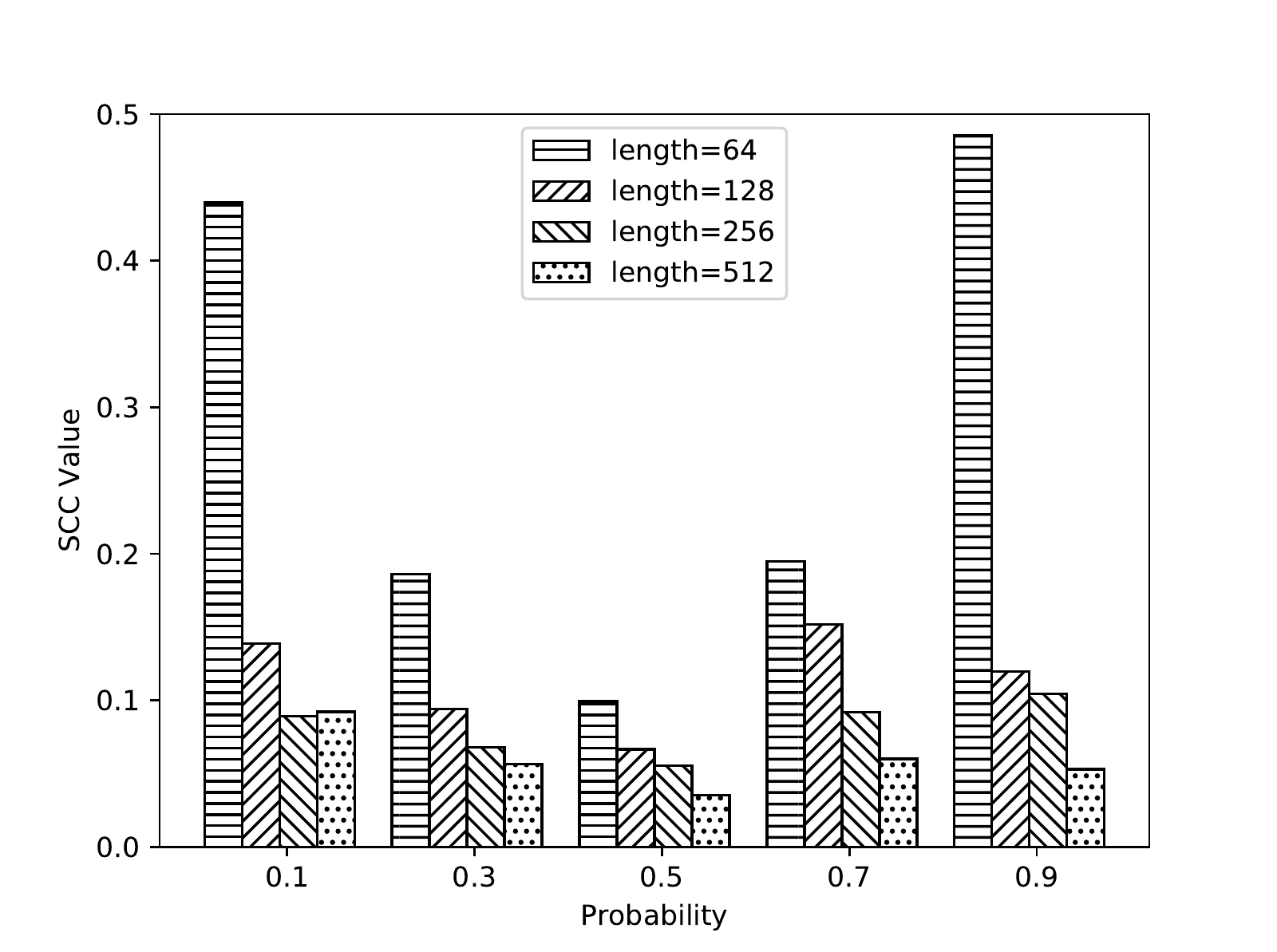}
    \caption{The self-$SCC$ measurement for probability $p \in \left[ { 0, 1} \right]$, i.e., evaluating the generated bitstreams for a particular probability while only $10\%,30\%,50\%,70\%,90\%$ are illustrated.}
    \label{fig:sameproscc}
\end{figure}

\begin{figure}[tb!]
    \centering
    \includegraphics[width=0.38\textwidth]{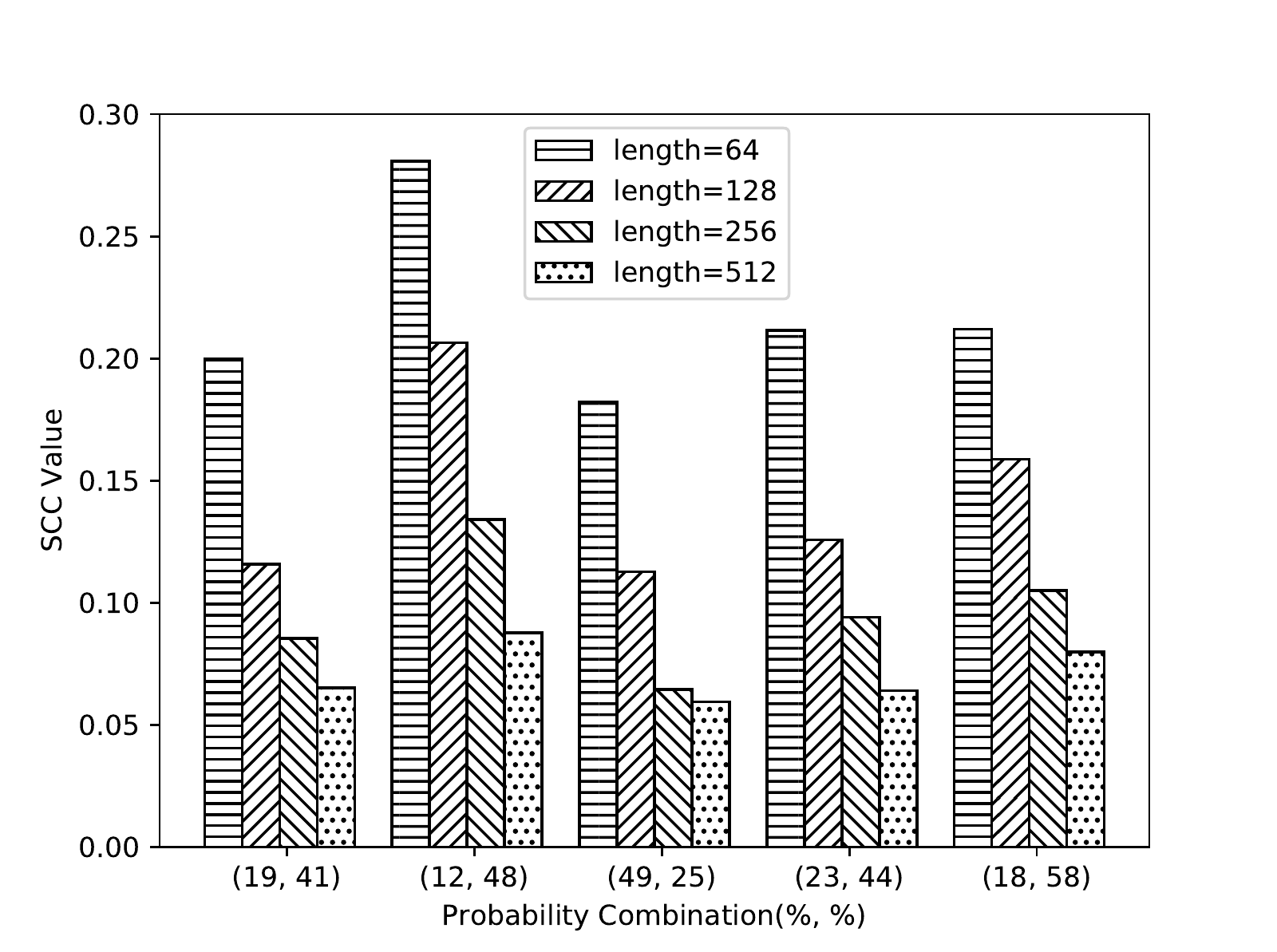}
    \caption{The cross-$SCC$ measurement for probability combinations, i.e., evaluating the generated bitstreams between different probabilities while only the combinations of ($19\%, 41\%$), ($12\%, 48\%$), ($49\%, 25\%$), ($23\%, 44\%$), ($18\%, 58\%$) are illustrated.}
    \label{fig:diffprosscc}
\end{figure}

As described above, stochastic computing usually requires a low correlation among different bitstreams. Many evaluation metrics of statistical correlation between different bitstreams have been proposed~\cite{choi2010survey}. The stochastic computing correlation ($SCC$) measurement~\cite{alaghi2013exploiting} is adopted in our work, which is particularly proposed for stochastic computing:
\begin{equation}
\small{SCC\left ( X_1, X_2 \right ) =} \begin{cases}
 &  \frac{ad-bc}{n \times min(a+b, a+c)-(a+b)(a+c)} \text{ if } ad>bc \\
 &  \frac{ad-bc}{(a+b)(a+c) - n \times max(a-d, 0)} \quad otherwise
\end{cases}
\label{eqn:scc}
\end{equation} where $X_1$ and $X_2$ are two stochastic bitstreams for measurement, $a$ is the number of `1's bit-overlapping between $X_1$ and $X_2$, $b$ is the number of bit-overlapping of `1's in $X_1$ and `0's in $X_2$, $c$ is the number of bit-overlapping of `0's in $X_1$ and `1's in $X_2$, $d$ is the number of `0's bit-overlapping between $X_1$ and $X_2$. From Eqn.~\eqref{eqn:scc}, $SCC \to +1$ if $X_1$ and $X_2$ have a maximum similarity; otherwise, $SCC \to -1$ if $X_1$ and $X_2$ are totally different. And consequently, we have $SCC \in \left[ { -1,1} \right]$. For a certain probability $p \in \left[ { 0, 1} \right]$, many bitstreams are generated to compute the $SCC$ absolute value, which is regarded as self-$SCC$ measurement. A  self-$SCC$ measurement sample is illustrated in Fig.~\ref{fig:sameproscc} with the bit length $n = 64, 128, 256, 512$. For measuring $SCC$ between different probabilities, two groups of bitstreams are generated to compute the $SCC$ absolute value, which is regarded as cross-$SCC$ measurement. A cross-$SCC$ measurement sample is demonstrated in Fig.~\ref{fig:diffprosscc} with the bit length $n = 64, 128, 256, 512$. As can be seen from Fig.~\ref{fig:sameproscc} and Fig.~\ref{fig:diffprosscc}, the $SCC$ values are relatively small so that they could satisfy the requirements of stochastic computing. And the $SCC$ value decreases when the bitstream length increases.

\begin{figure}[tb!]
    \centering
    \includegraphics[width=0.35\textwidth]{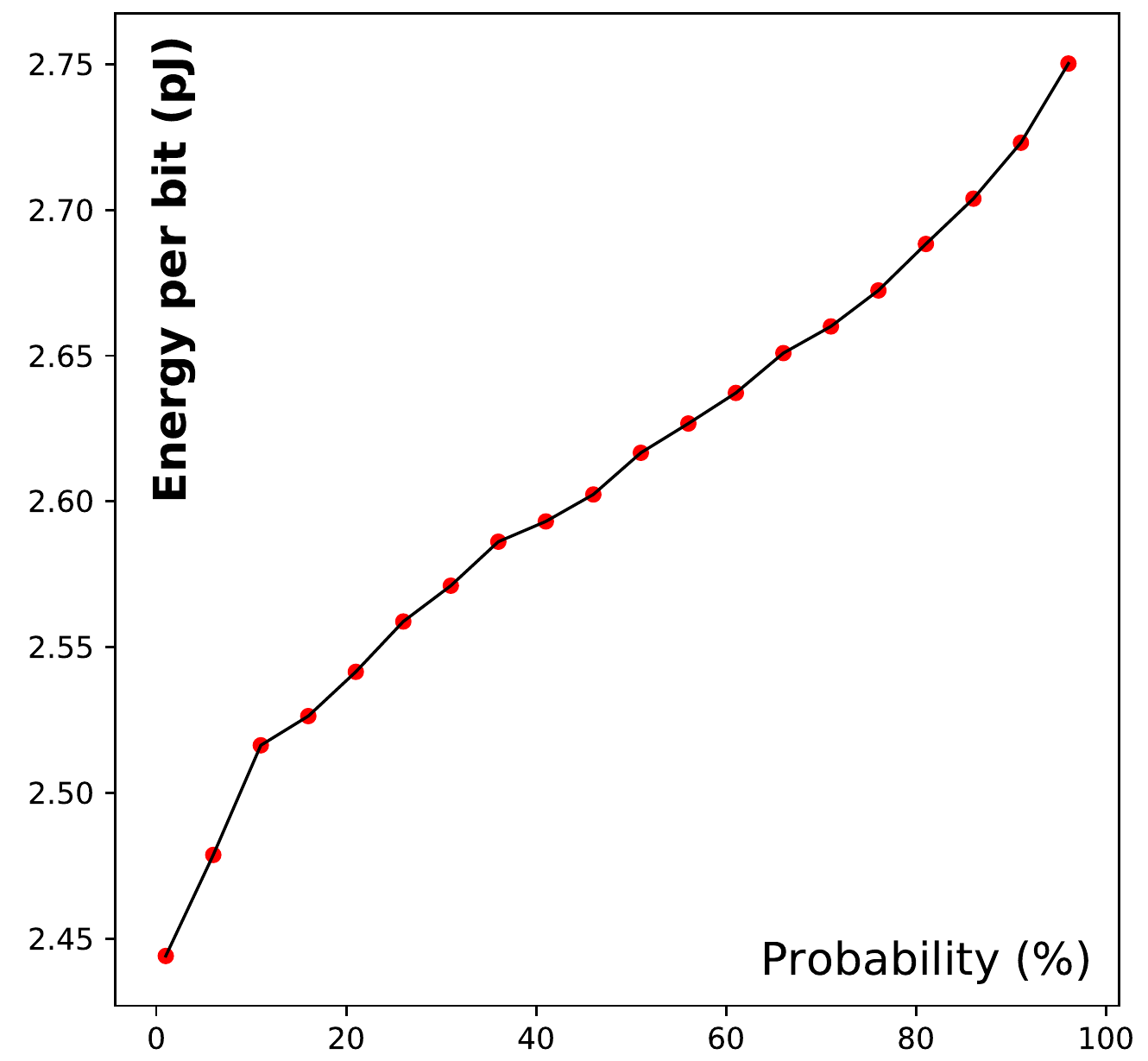}
    \caption{Energy consumption for generating bitstreams of probability $p \in \left[ { 0, 1} \right]$.}
    \label{fig:sbg_energy}
\end{figure}

There are $93$ CMOS transistors, one resistor and 5 MTJs in the proposed self-control SBG circuits which will be adopted to analyze the occupied chip area. Also the energy consumption for generating bitstreams of probability $p \in \left[ { 0, 1} \right]$ is demonstrated in Fig.~\ref{fig:sbg_energy}, from which we can see the larger probability usually requires more energy consumption.

\begin{table}
\centering
\caption{Simulation results of probability-voltage relationship under the certain process variation.}
\label{Table:pvrelationwithprocessvariation}
\begin{tabular}{c|ccc}
\specialrule{0.8pt}{0pt}{0pt}
Bitstream Length &         64 &        128 &        256 \\
\hline
Max Error &     0.1295 &   0.0949 &    0.0733 \\
Avg. Error &    0.0460 &    0.0336 &   0.0269 \\
\specialrule{0.8pt}{0pt}{0pt}
\end{tabular}
\end{table}

\subsubsection{Process Variation} \label{Section:sbg:evaluation:pv}
MTJ switching behavior is deeply impacted by the process variation such as MTJ geometric variation and initial magnetization angle variation~\cite{nigam2011delivering,emre2012enhancing}. Variation in surface area ($A$) and tunneling oxide thickness ($t_{ox}$) are the main causes behind the resistance change in MTJ material because $R_{MTJ} \propto (1/A) \cdot e^{t_{ox}}$. Assuming that the variation of $A$ and $t_{ox}$ follows Gaussian distribution with a standard deviations of 5\% and 2\% of their mean value, respectively~\cite{Li2008modeling,zhang2011stt}, the sensitivity of the relationship between required probability and applied voltage is evaluated and shown in Table~\ref{Table:pvrelationwithprocessvariation}. The accuracy could be improved by increasing the bitstream length.

 \iffalse

 \bibliography{../ref/Bayesian}

 \fi

\section {Applications}\label{Section:experiments}

As demonstrated in Fig.~\ref{fig:spinbis}, the stochastic computing architectures are determined according to the specified applications. A device-to-architecture level evaluation framework is illustrated for SPINBIS and a typical application is demonstrated as a case study in this section.

\subsection{Evaluation Framework} \label{Section:experiments:framework}

\begin{figure}[tb!]
    \centering
    \includegraphics[width=0.45\textwidth]{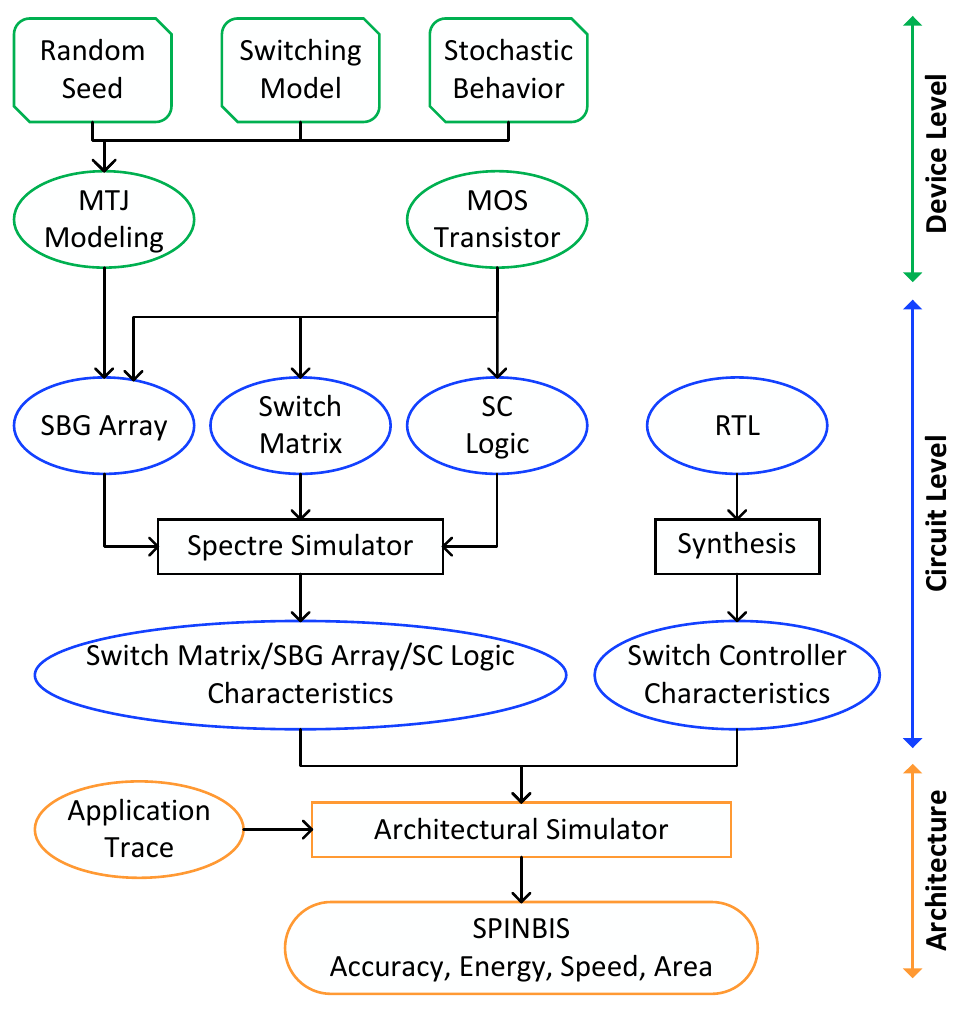}
    \caption{Evaluation framework of SPINBIS.}
    \label{fig:framework}
\end{figure}

SPINBIS is implemented by hybrid CMOS/MTJ technologies with three design hierarchies: device, circuit and architecture levels as shown in Fig~\ref{fig:framework}. The hybrid CMOS/MTJ circuits are simulated by Cadence Spectre simulator while the MTJ model is written by Verilog-A language. The dynamic switching of MTJ device is realized with two regimes of Sun model \cite{worledge2011spin} and Neel-Brown model \cite{heindl2011validity}. The stochastic MTJ switching behaviors are modeled by \cite{devolder2008single}. In order to reduce the correlation of bitstreams generated by different SBG circuits, the random seed is configured as different for different MTJ instances as described in Section~\ref{Section:sbg:evaluation}. With the circuit simulation results, the SBG array, switch matrix and stochastic computing logics are abstracted as behavioral blocks by performing characterizations. Meanwhile, the RTL implementation of switch controller is synthesized by Synopsys Design Compiler with $45$ $nm$ FreePDK library. After performing the characterization of switch controller, an architectural level simulation is carried out according to the specified application trace. Finally, the evaluation results of SPINBIS are obtained in terms of inference accuracy, energy efficiency, inference speed and design area.

\begin{figure*}[tb!]
    \centering
    \hspace{-2mm}
    \includegraphics[scale=1.2]{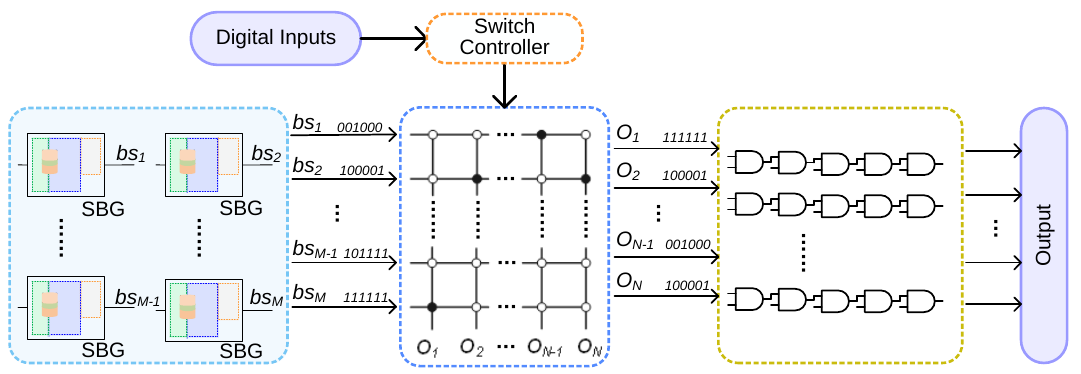}
    \caption{SPINBIS diagram of target locating problem.}
    \label{fig:df_system}
\end{figure*}

\begin{figure}[tb!]
    \centering
    \hspace{-2mm}
    \includegraphics[width=0.48\textwidth]{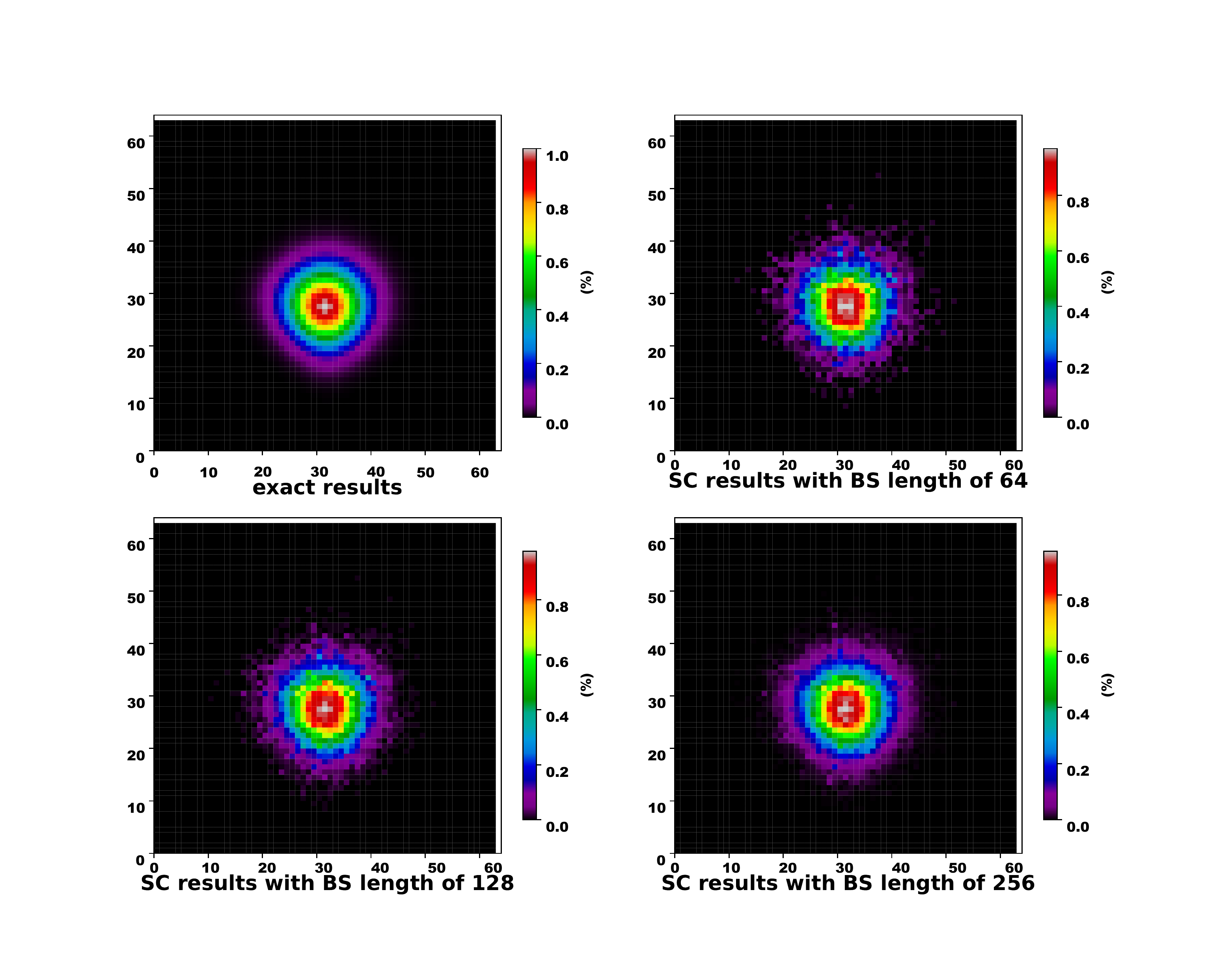}
    \caption{Sensor fusion results of SPINBIS for target locating problem on 64 $\times$ 64 grid compared with exact solutions.}
    \label{fig:df_results}
\end{figure}

\subsection {Case Study: Data Fusion for Target Locating}\label{Section:experiments:df}

Data fusion aims to achieve more consistent, accurate, and useful information by integrating multiple data sources instead of by any individual data source.  In this section, a simple data fusion example is demonstrated and the corresponding Bayesian inference procedures are also studied.

\subsubsection{Problem definition and Bayesian inference algorithm}

Sensor fusion aims to determine a target location by multiple sensors~\cite{coninx2016Bayesian}. Assuming that there are $3$ sensors on a $2D$ plane while the width and length of $2D$ plane is $64$ and sensors are located at $\left( {0,0} \right)$, $\left( {0,32} \right)$, $\left( {32,0} \right)$. Each sensor has 2 data channels: distance ($d$) and bearing ($b$). The measured data $(d_1, b_1, d_2, b_2, d_3, b_3)$ from $3$ sensors with $2$ channels are utilized to calculated the target location $(x^\star,y^\star)$. In data fusion application, the probability value that target object locates at one position of the plane is calculated based on the sensor data. The position with the largest probability value is considered to be the position that object target is located at.

Based on the observed data $(d_1, b_1, d_2, b_2, d_3, b_3)$, the probability of target object located on $(x,y)$ is denoted as $p(x,y|d_1,b_1,d_2,b_2,d_3,b_3)$ and could be calculated based on Bayes' theory:
\begin{equation}
\label{Equation:pro_simp}
p(x,y|d_1,b_1,d_2,b_2,d_3,b_3)
\propto  p(x,y ) * \prod_{i} { p(d_i|x, y)p(b_i|x,y) }
\end{equation} where $p(x,y)$ is denoted as prior probability, and $p(d_i|x,y),p(b_i|x,y)$ are known as evidence or likelihood information. Since the target may locate at any position, the prior probability $p(x,y)$ is the same for any position. Hence, $p(x,y)$ is ignored in the following Bayesian inference system. $p(d_i|x,y)$ means the probability that the $i$-th sensor return the distance value of $d_i$ if the target object is located at position $(x, y)$. The meaning of $p(b_i|x,y)$ is similar to that of $p(d_i|x,y)$. The value of $p(d_i|x,y)$ and $p(b_i|x,y)$ is calculated by \eqref{Equation:GaussianDistribution}.

\begin{equation}
\label{Equation:GaussianDistribution}
\begin{array}{*{20}{c}}
p(d_i|x, y) = \frac { 1 }{ \sqrt { 2\pi  } {\sigma}_{i}^{d}  } \cdot { e }^{ -\frac { { \left(d\left(x, y\right)- {\mu}_{i}^{d} \right) }^{ 2 } }{ 2 \left({\sigma}_{i}^{d}  \right)^{ 2 } }  } \\
p(b_i|x, y) = \frac { 1 }{ \sqrt { 2\pi  } {\sigma}_{i}^{b}  } \cdot { e }^{ -\frac { { \left(b\left(x, y\right)- {\mu}_{i}^{b} \right) }^{ 2 } }{ 2{ \left({\sigma}_{i}^{b}  \right)}^{ 2 } }  }
\end{array}
\end{equation} where $d(x, y)$ is the Euclidian distance between position $(x, y)$ and the $i$-th sensor, ${\mu}_{i}^{d}$ is the distance data provided by the $i$-th sensor, ${\sigma}_{i}^{d} = 5 + {\mu}_{i}^{d}/10$. $b(x, y)$ is the viewing angle from the $i$-th sensor to position $(x,y)$,  ${\mu}_{i}^{b}$ is the bear data provided by the $i$-th sensor, ${\sigma}_{i}^{b}$ is set as 14.0626 degree.

\subsubsection{Bayesian inference system}

From Eqn.~\eqref{Equation:pro_simp}, the Bayesian inference is calculated by the product of a series of conditional probabilities, which could be realized by performing stochastic computing with \texttt{AND} gates and stochastic bitstreams. Given any $2$ positions $(x_1, y_1)$ and $(x_2, y_2)$, the calculations of $p(x_1 ,y_1|d_1,b_1,d_2,b_2,d_3,b_3)$ and $p(x_2 ,y_2|d_1,b_1,d_2,b_2,d_3,b_3)$ could be finished in parallel since they are independent with each other.

The SPINBIS architectures are reformulated as Fig.~\ref{fig:df_system} for sensor fusion applications. For each probability calculation, it requires $5$ \texttt{AND} gates to perform multiplications for $6$ conditional probabilities. The SBG sharing and terminal clustering strategies are utilized to reduce the required scale of SBG array and switch matrix. The $2D$ plane is partitioned as $64 \times 64$ and $32 \times 32$ grids for target locating problem. The finer grid partition usually achieves more accurate locating results. Table~\ref{Table:scale} shows the scale of SBG array and switch matrix for different grid size. The meaning of symbol $T$, $N$, $M$ and $N'$ in Table~\ref{Table:scale} have been described in Section~\ref{Section:spinbis:discussion}. For $64 \times 64$ grid size problem, the energy consumption and transistor utilization of SBG array and switch matrix are $1.3\%$ and $80\%$ of that in~\cite{jia2017spintronics}, respectively. For $32 \times 32$ grid size problem, the energy consumption is $5.2\%$ of that in~\cite{jia2017spintronics} while the transistor utilization is $1.64\times$ of that in~\cite{jia2017spintronics}.

\begin{table}
\centering
\caption{Transistor utilizations of SBG array and switch matrix for different grid size, where ${\cal{K}}_{energy} = \frac {M} {N}$ and ${\cal{K}}_{cmos} = \frac {T * M + M * N'} {T * N}$ indicates the improvement on energy and area efficiency, respectively.}
\label{Table:scale}
\begin{tabular}{c|c|c|c|c|c|c}
\specialrule{0.8pt}{0pt}{0pt}
 Grid Size &          $T$ &          $N$ &          $M$ &         $N'$ &   ${\cal{K}}_{energy}$ &     ${\cal{K}}_{cmos}$ \\
\hline
     $32 \times 32$ &         92 &       6144 &        320 &       2817 &      0.052 &   1.64 \\

     $64 \times 64$ &         92 &      24576 &        320 &       5557 &      0.013 &   0.79 \\
\specialrule{0.8pt}{0pt}{0pt}
\end{tabular}
\end{table}

% T*N = 2260,992
% T*M = 29940
% M*N' = 1778,240

\subsubsection{Simulation results}

\begin{figure}[tb!]
    \centering
    \hspace{-2mm}
    \includegraphics[width=0.4\textwidth]{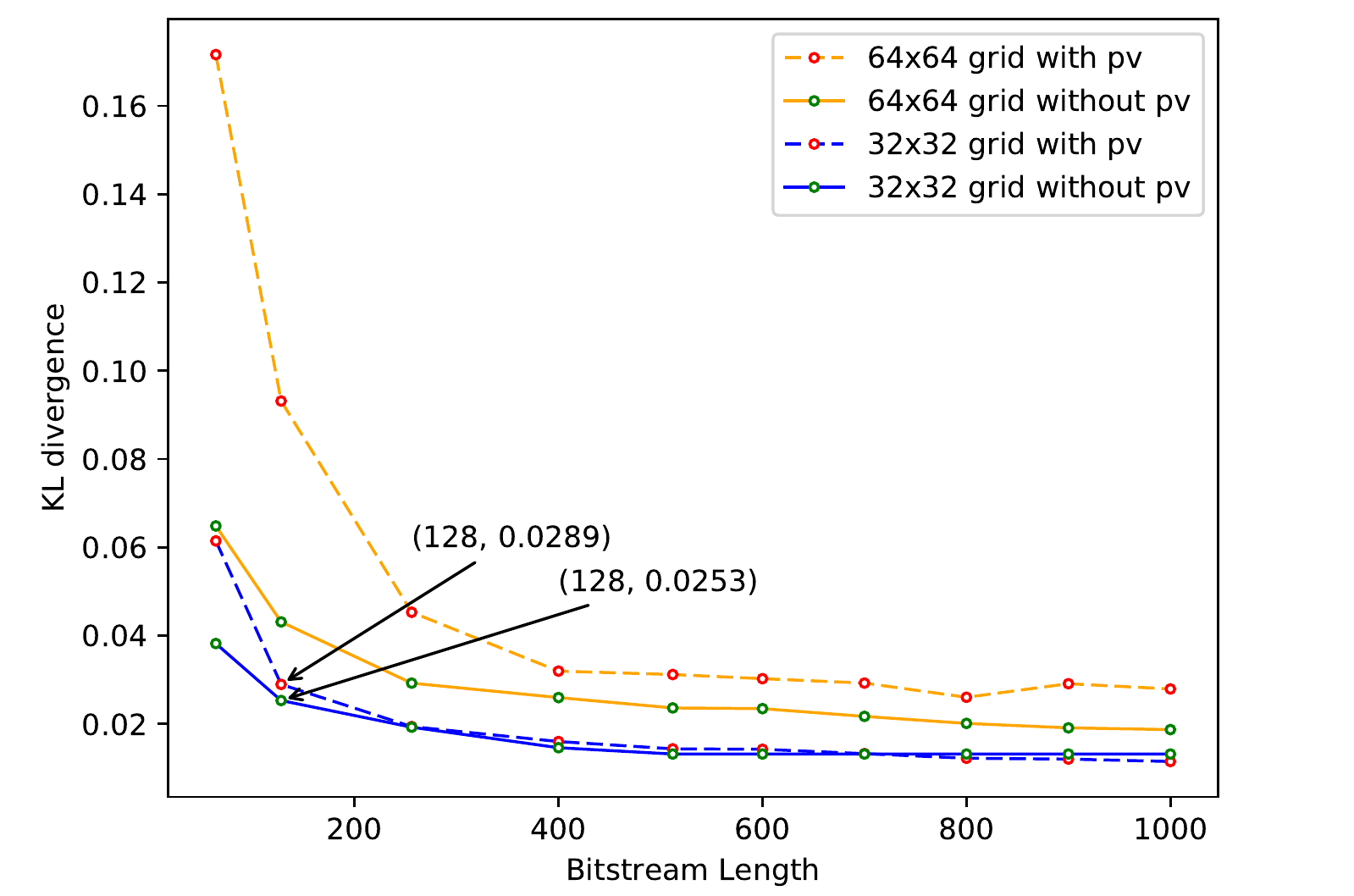}
    \caption{KL divergence analysis of SPINBIS for target locating problem.}
    \label{fig:kl_diver}
\end{figure}

The obtained data from sensors is represented as bitstreams with length of $64$, $128$ and $256$ for stochastic computing. The fusion results on $64 \times 64$ grid are shown as heat maps in Fig.~\ref{fig:df_results} and compared with exact result. The bitstream with larger length usually has a better inference accuracy. Meanwhile, Kullback-Leibler (KL) divergence is further introduced to measure the differences between stochastic inference results and exact solutions as shown in Fig.~\ref{fig:kl_diver}.
The dashed yellow line and blue line represent the KL divergence value under the specified process variations for $64 \times 64$ grid and $32 \times 32$ grid, respectively. While the solid yellow line and blue line represent the KL divergence value without process variations. For the same KL divergence value, the length of bitstream without process variation is usually larger than that without process variation but still smaller than the length of work~\cite{coninx2016Bayesian}. As reported in previous work~\cite{coninx2016Bayesian}, the sensor fusion on $32 \times 32$ grid for \textbf{$\bm{10^4}$ cycles} could obtain a KL divergence of $0.029$. However, SPINBIS only need about \textbf{128 cycles} to achieve such a KL divergence as shown in Fig.~\ref{fig:kl_diver} even with the consideration of process variation. In summary, these advantages benefit from the high accuracy and low correlation bitstreams generated by the MTJ based SBG array.

\subsubsection{Performance}

\begin{table}
\caption{Comparison between stochastic computing method and 8-bit fixed point binary implementation on FPGA.}
\label{Table:SCvsBinary}
\begin{tabular}{c|c|c|c|c|c}
\specialrule{0.8pt}{0pt}{0pt}
\multirow{2}{*}{Method} &  Bitstream & KL         & \multirow{2}{*}{Energy} & \multicolumn{2}{c}{Utilization} \\
% \cmidrule(lr){3-4}
\cline{5-6}
                        &  Length    & Divergence &                         &         LUT & FF \\
\hline
\multirow{4}{*}{SC} &      64 &     0.051 &          0.66$\mu J$ &     \multirow{4}{*}{9316} &    \multirow{4}{*}{17608} \\
                    &     128 &     0.037 &          1.32$\mu J$ &                           &                           \\
                    &     200 &     0.031 &          2.06$\mu J$ &                           &                           \\
                    &     256 &     0.021 &          2.64$\mu J$ &                           &                           \\
                    &     512 &     0.014 &          5.28$\mu J$ &                           &                           \\
\hline
    Binary          &      - &         - &           1.99$\mu J$ &       234496               &     253952 \\
\specialrule{0.8pt}{0pt}{0pt}
\end{tabular}
\end{table}

The performance of SPINBIS with the considerations of process variations is compared with FPGA~\cite{coninx2016Bayesian} and MTJ~\cite{jia2017spintronics} based approaches. The sensor fusion problem is evaluated by these approaches on the 32$\times$32 grid.

Firstly, the stochastic computing method is compared with 8-bit fixed point binary implementation on the same FPGA platform of Xilinx Zynq 7020. The stochastic computing method is referred to~\cite{coninx2016Bayesian} but re-implemented by ourselves for the sake of fairness and clarity. In Table~\ref{Table:SCvsBinary}, stochastic computing results are illustrated with different bitstream length. As shown in Table~\ref{Table:SCvsBinary}, longer bitstream realization could obtain a lower KL divergence (better accuracy) but requires more energy consumption. Once the bitstream length is larger than about 200, stochastic computing method consumes more energy than 8-bit fixed point binary implementation. Additionally, the resources utilization of stochastic computing approach is much lower than binary implementation. In fact, stochastic computing method provides a trade-off between energy consumption and inference accuracy. Hence, stochastic computing is very promising for fault-tolerant embedded applications which require higher area efficiency. Then the comparison of stochastic computing results of different approaches are illustrated in Table~\ref{Table:energy}. All of the inference approaches are required to satisfy the requirement of KL divergence less than $0.029$. As seen from Table~\ref{Table:energy}, the energy efficiency of MTJ-based approach~\cite{jia2017spintronics} is significantly improved than FPGA approach~\cite{coninx2016Bayesian}. Furthermore, SPINBIS achieves better energy efficiency and inference speed compared with MTJ~\cite{jia2017spintronics} and FPGA~\cite{coninx2016Bayesian} approaches and bring about $45\%$ design area overhead compared with MTJ-based approach~\cite{jia2017spintronics}.

\begin{table}
\caption{SPINBIS performance comparison with other methods with the requirement of KL divergence less than 0.029 on $32 \times 32$ grid, where $E_{cyc}$ is energy consumption of each cycle, $T_{cyc}$ is the duration time of each cycle, $N_{cyc}$ is the total cycle count, $E_{tot}$ is the total energy consumption for all cycles, $T_{tot}$ is the total inference time, $N_{cmos}$ is the number of utilized CMOS transistors.}
\label{Table:energy}
\begin{tabular}{l|r|r|r|r|r|c}
    \specialrule{0.8pt}{0pt}{0pt}
    \multirow{2}{*}{Method}     & $E_{cyc}$ & $T_{cyc}$ & \multirow{2}{*}{$N_{cyc}$} & $E_{tot}$ & $T_{tot}$ & $N_{cmos}$ \\
         & ($nJ$)      & ($ns$)      &     & ($\mu J$)   & ($\mu s$)   & ($\times 10^3$) \\
    \hline
    FPGA                          & $10.3$ & $10$ & 256 & $2.64$ & $2.56$   & -  \\
    MTJ~\cite{jia2017spintronics} & $4.58$ & $40$ & 256 & $1.17$ & $10.24$  & $\approx 830$  \\
    SPINBIS                       & $0.78$ & $10$ & 128 & $0.10$ & $1.28$   & $\approx 1200$  \\
    \specialrule{0.8pt}{0pt}{0pt}
\end{tabular}
\end{table}

\section{Conclusion}\label{Section:Conclusion}

Spintronic device is a promising technology because of its low power, high speed, infinite endurance and easy integration with CMOS circuit. In this paper, the inherent stochastic behavior of MTJ is utilized to build the stochastic bitstream generator which is critical for Bayesian inference system. A state-aware self-control strategy is proposed to improve the energy efficiency and speed of SBG circuit. The SBG sharing strategy and terminal clustering strategy are further proposed in SPINBIS to reduce the energy consumption and design area overhead. A device-to-architecture level framework is demonstrated to evaluate the performance of SPINBIS and a typical application is demonstrated as a case study. Experimental results on data fusion applications demonstrate that SPINBIS could improve the energy efficiency about $12\times$ than MTJ-based approach with $45\%$ design area overhead and about $26\times$ than FPGA-based approach.

In the future, we will carry on our research on the following aspects. Firstly, the probability and voltage relation is not very smooth. It is necessary to improve the stability of the proposed SBG. Secondly, the adopted switch matrix could still have a congestion problem even if the scale is reduced from $M \times N$ to $M \times N'$. Further reduction of the scale of SPINBIS is also a desirable research point.

%{
%\bibliographystyle{IEEEtran}
%\bibliography{./ref/Top_sim,./ref/Bayesian}
%}


\vspace{-5mm}

\begin{IEEEbiography}[{\includegraphics[width=1in,height=1in,clip,keepaspectratio]{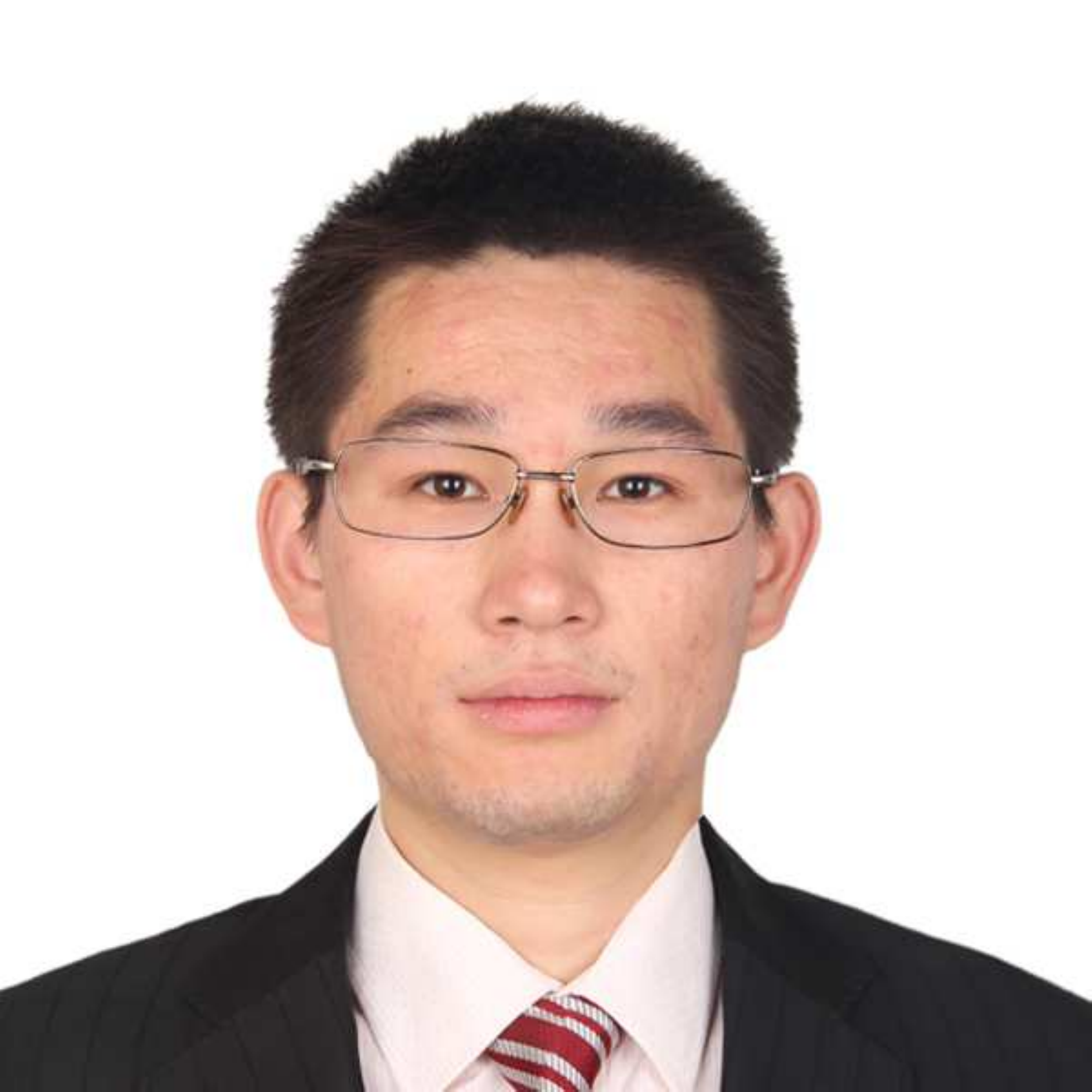}}]{Xiaotao Jia}

(S'13-M'17) received the B.S. degree in mathematics from Beijing Jiao Tong University, Beijing, China, in 2011, and the Ph.D. degree in computer science and technology from Tsinghua University, Beijing, China, in 2016.

He is currently a post-doctoral researcher with the Fert Beijing Research Institute in Beihang University, Beijing, China. His current research interests include spintronic circuits and Bayesian learning systems.

\end{IEEEbiography}

\vspace{-5mm}

\begin{IEEEbiography}[{\includegraphics[width=1in,height=1.25in,clip,keepaspectratio]{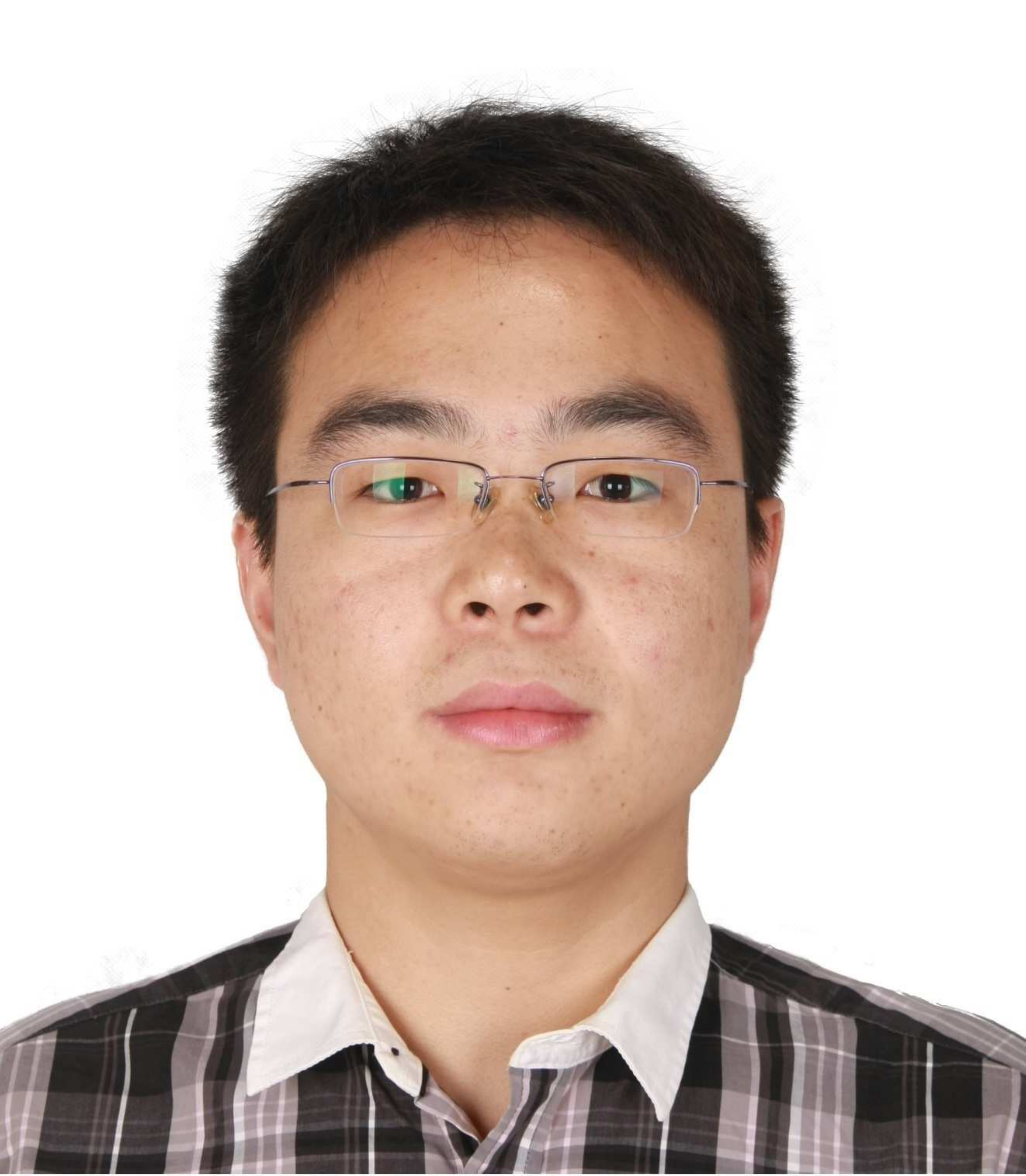}}]{Jianlei Yang}

(S'12-M'16) received the B.S. degree in microelectronics from Xidian University, Xi'an, China, in 2009, and the Ph.D. degree in computer science and technology with Tsinghua University, Beijing, China, in 2014.

He joined Beihang University, Beijing, China, in 2016, where he is currently an Associate Professor with the School of Computer Science and Engineering. From 2014 to 2016, he was a post-doctoral researcher with the Department of Electrical and Computer Engineering, University of Pittsburgh, Pittsburgh, Pennsylvania, United States. From 2013 to 2014, he was a research intern at Intel Labs China, Intel Corporation. His current research interests include spintronics and neuromorphic computing systems.

Dr. Yang was the recipient of the first place on TAU Power Grid Simulation Contest in 2011, and the second place on TAU Power Grid Transient Simulation Contest in 2012. He was a recipient of IEEE ICCD Best Paper Award in 2013, IEEE ICESS Best Paper Award in 2017, and ACM GLSVLSI Best Paper Nomination in 2015.

\end{IEEEbiography}

\vspace{-5mm}

\begin{IEEEbiography}[{\includegraphics[width=1in,height=1.25in,clip,keepaspectratio]{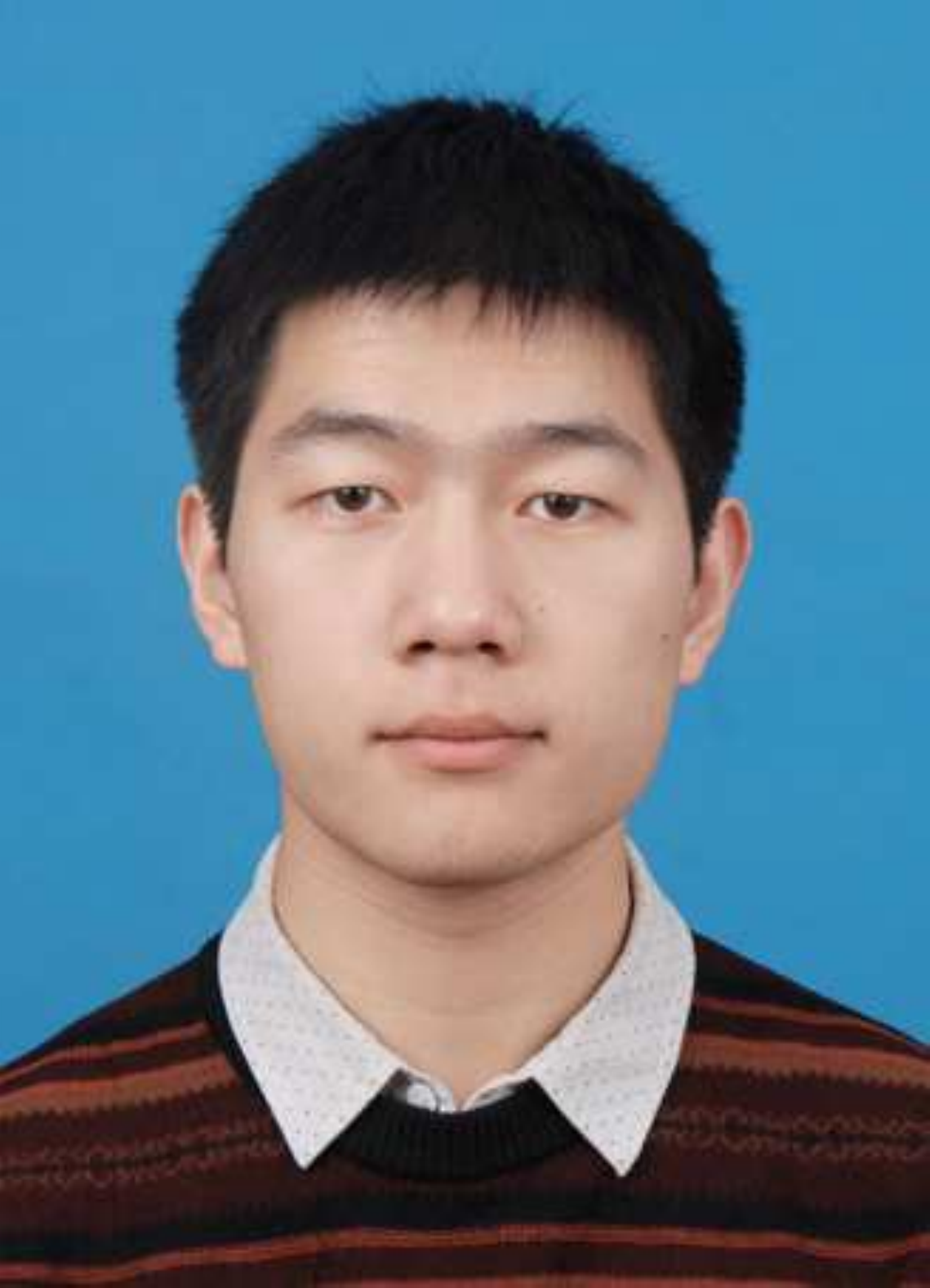}}]{Pengcheng Dai}

received the B.S. degree in electronic engineering from Beihang University, Beijing, China, in 2017. He is currently a graduate student in School of Electronic and Information Engineering, Beihang University, Beijing, China. His research interests include computing architectures for deep learning and machine vision.

\end{IEEEbiography}

\vspace{-5mm}

\begin{IEEEbiography}[{\includegraphics[width=1in,height=1.25in,clip,keepaspectratio]{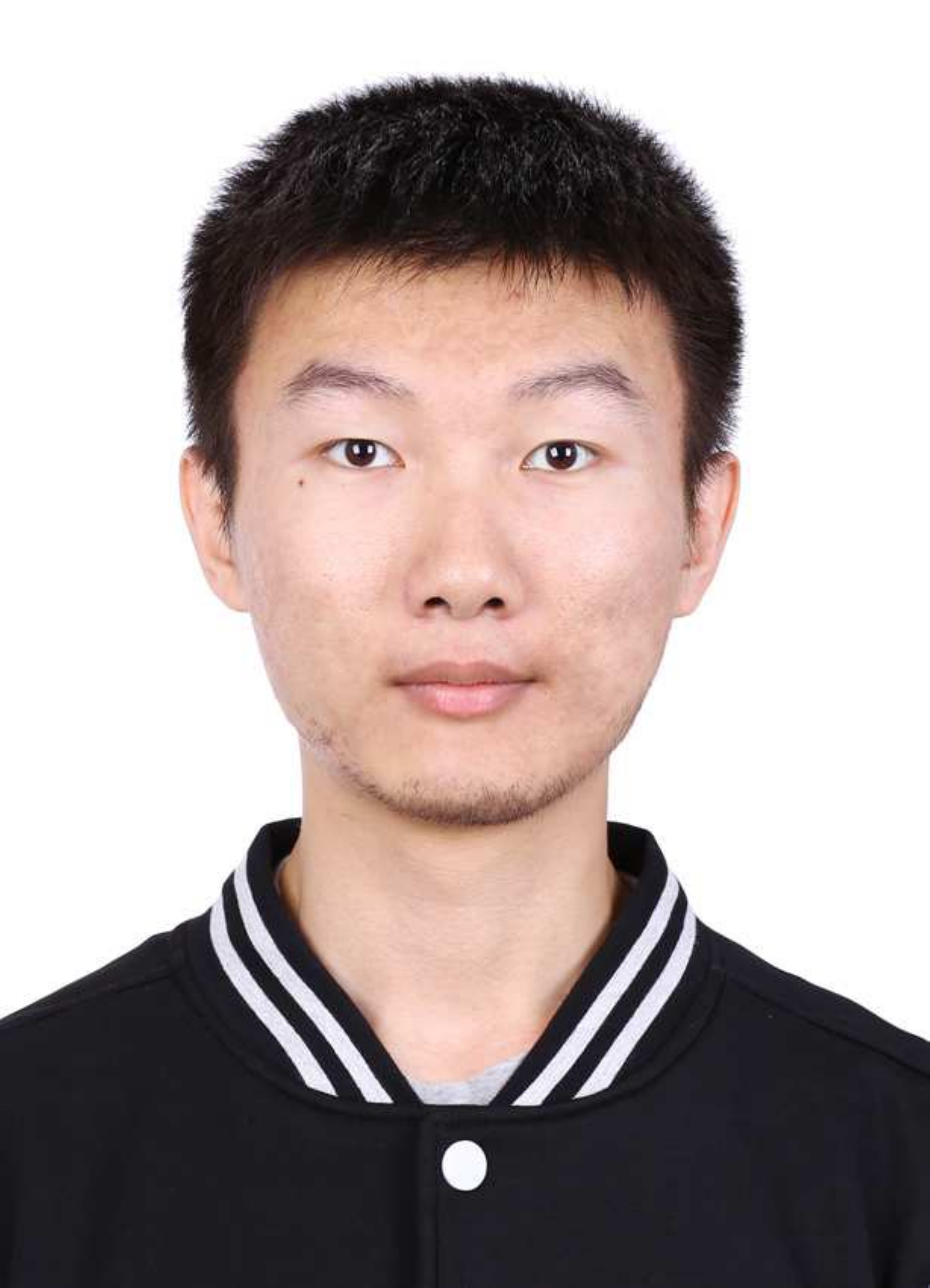}}]{Runze Liu}

received the B.S. degree in School of Computer Science and Engineering, from Beihang University, Beijing, China, in 2018. He is currently a graduate student in University of South California, CA, USA. His research interests include computing architectures for deep learning and machine vision.

\end{IEEEbiography}

\vspace{-5mm}

\begin{IEEEbiography}[{\includegraphics[width=1in,height=1.25in,clip,keepaspectratio]{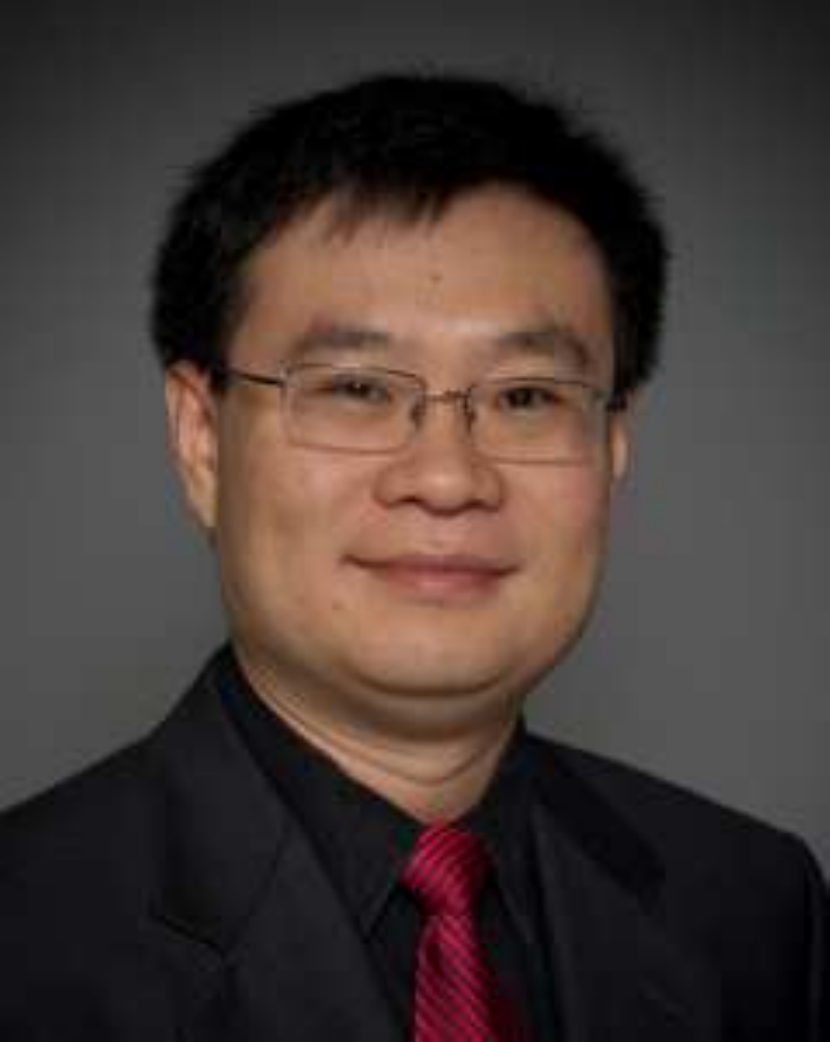}}]{Yiran Chen}

(M'04-SM'16-F'18) received B.S and M.S. from Tsinghua University and Ph.D. from Purdue University in 2005. After five years in industry, he joined University of Pittsburgh in 2010 as Assistant Professor and then promoted to Associate Professor with tenure in 2014, held Bicentennial Alumni Faculty Fellow. He now is a tenured Associate Professor of the Department of Electrical and Computer Engineering at Duke University and serving as the co-director of Duke Center for Evolutionary Intelligence (CEI), focusing on the research of new memory and storage systems, machine learning and neuromorphic computing, and mobile computing systems. Dr. Chen has published one book and more than 300 technical publications and has been granted 93 US patents. He is the associate editor of IEEE TNNLS, IEEE TCAD, IEEE D\&T, IEEE ESL, ACM JETC, ACM TCPS, and served on the technical and organization committees of more than 40 international conferences. He received 6 best paper awards and 14 best paper nominations from international conferences. He is the recipient of NSF CAREER award and ACM SIGDA outstanding new faculty award. He is the Fellow of IEEE.

\end{IEEEbiography}

\vspace{-5mm}

\begin{IEEEbiography}[{\includegraphics[width=1in,height=1.25in,clip,keepaspectratio]{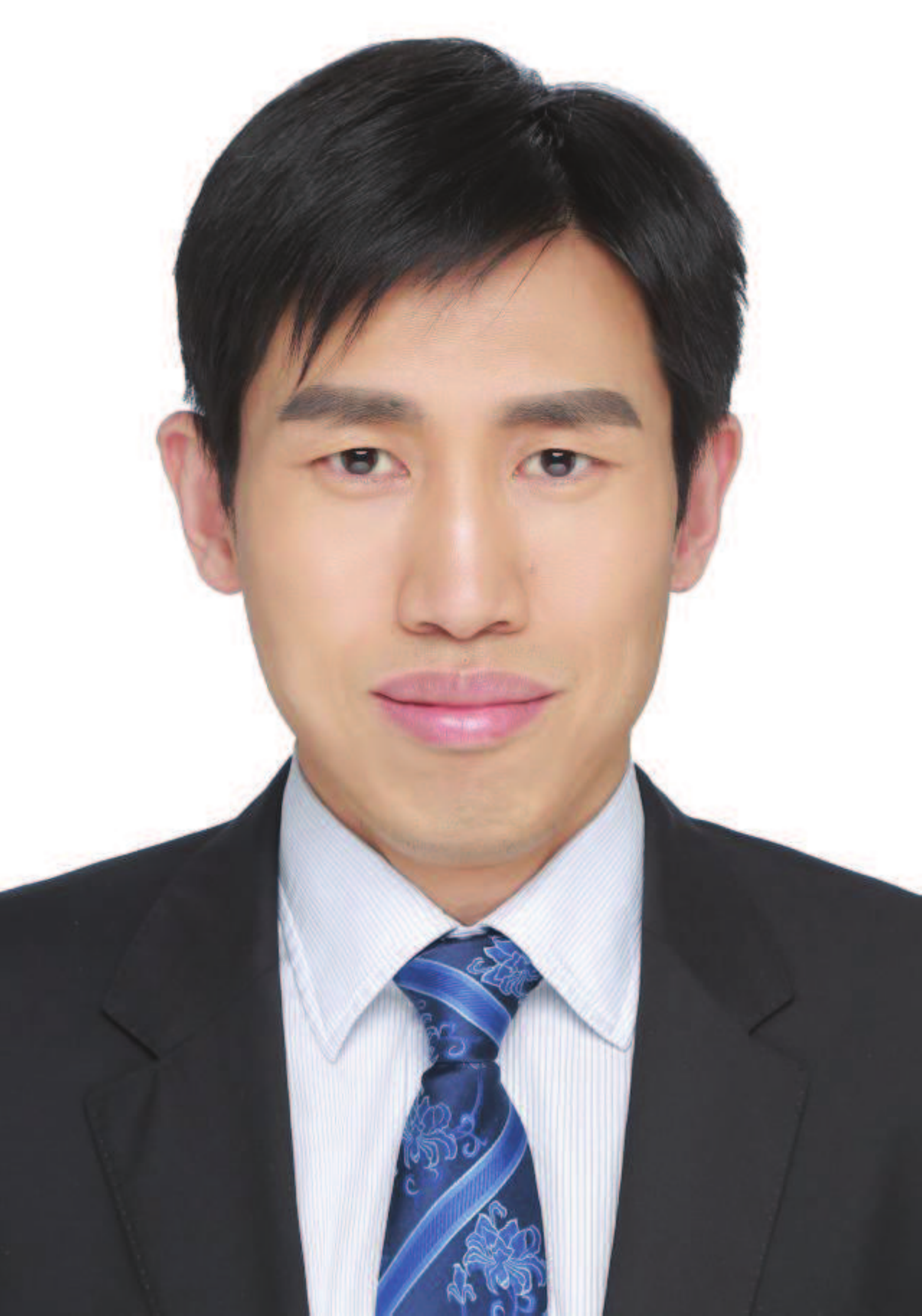}}]{Weisheng Zhao}

(M'06-SM'14-F'19) received the Ph.D. degree in physics from University of Paris Sud, Paris, France, in 2007.

He worked as a Research Associate at the CEA's embedded computing laboratory, France, from 2007 to 2009, and at the French national research center (CNRS), France, as a tenured scientist from 2009 to 2014 where he led the spintronics integration group. Now he is a professor and director of Fert Beijing Research Institute in Beihang University, Beijing, China. He has authored or coauthored 2 books, more than 200 scientific papers in the leading journals such as Nature Communications, Advanced Materials, Proceedings of the IEEE and he also holds 4 international patents and more than 50 Chinese patents. He is the Fellow of IEEE.

Prof. Zhao is the associated editor of {\sc{IEEE Transactions on Nanotechnology}} and {\sc{IET Electronics Letters}}.

\end{IEEEbiography}

\end{document}